\PassOptionsToPackage{unicode}{hyperref}
\PassOptionsToPackage{hyphens}{url}
\PassOptionsToPackage{dvipsnames,svgnames,x11names}{xcolor}
\documentclass[
  12pt]{article}

\usepackage{amsmath,amssymb}
\usepackage[T1]{fontenc}
\usepackage[utf8]{inputenc}
\usepackage{textcomp}
\usepackage{lmodern} 

\usepackage{lmodern}
\IfFileExists{upquote.sty}{\usepackage{upquote}}{}
\IfFileExists{microtype.sty}{
  \usepackage[]{microtype}
  \UseMicrotypeSet[protrusion]{basicmath} 
}{}
\makeatletter
\@ifundefined{KOMAClassName}{
  \IfFileExists{parskip.sty}{%
    \usepackage{parskip}
  }{
    \setlength{\parindent}{0pt}
    \setlength{\parskip}{6pt plus 2pt minus 1pt}}
}{
  \KOMAoptions{parskip=half}}
\makeatother
\usepackage{xcolor}
\makeatletter
\ifx\paragraph\undefined\else
  \let\oldparagraph\paragraph
  \renewcommand{\paragraph}{
    \@ifstar
      \xxxParagraphStar
      \xxxParagraphNoStar
  }
  \newcommand{\xxxParagraphStar}[1]{\oldparagraph*{#1}\mbox{}}
  \newcommand{\xxxParagraphNoStar}[1]{\oldparagraph{#1}\mbox{}}
\fi
\ifx\subparagraph\undefined\else
  \let\oldsubparagraph\subparagraph
  \renewcommand{\subparagraph}{
    \@ifstar
      \xxxSubParagraphStar
      \xxxSubParagraphNoStar
  }
  \newcommand{\xxxSubParagraphStar}[1]{\oldsubparagraph*{#1}\mbox{}}
  \newcommand{\xxxSubParagraphNoStar}[1]{\oldsubparagraph{#1}\mbox{}}
\fi
\makeatother

\usepackage{longtable,booktabs,array}
\usepackage{calc} 
\usepackage{etoolbox}
\makeatletter
\patchcmd\longtable{\par}{\if@noskipsec\mbox{}\fi\par}{}{}
\makeatother
\IfFileExists{footnotehyper.sty}{\usepackage{footnotehyper}}{\usepackage{footnote}}
\makesavenoteenv{longtable}
\usepackage{graphicx}
\usepackage{authblk}
\makeatletter
\def\maxwidth{\ifdim\Gin@nat@width>\linewidth\linewidth\else\Gin@nat@width\fi}
\def\maxheight{\ifdim\Gin@nat@height>\textheight\textheight\else\Gin@nat@height\fi}
\makeatother
\setkeys{Gin}{width=\maxwidth,height=\maxheight,keepaspectratio}
\makeatletter
\def\fps@figure{htbp}
\makeatother

\makeatletter
\@ifpackageloaded{caption}{}{\usepackage{caption}}
\AtBeginDocument{%
\ifdefined\contentsname
  \renewcommand*\contentsname{Table of contents}
\else
  \newcommand\contentsname{Table of contents}
\fi
\ifdefined\listfigurename
  \renewcommand*\listfigurename{List of Figures}
\else
  \newcommand\listfigurename{List of Figures}
\fi
\ifdefined\listtablename
  \renewcommand*\listtablename{List of Tables}
\else
  \newcommand\listtablename{List of Tables}
\fi
\ifdefined\figurename
  \renewcommand*\figurename{Figure}
\else
  \newcommand\figurename{Figure}
\fi
\ifdefined\tablename
  \renewcommand*\tablename{Table}
\else
  \newcommand\tablename{Table}
\fi
}
\@ifpackageloaded{float}{}{\usepackage{float}}
\floatstyle{ruled}
\@ifundefined{c@chapter}{\newfloat{codelisting}{h}{lop}}{\newfloat{codelisting}{h}{lop}[chapter]}
\floatname{codelisting}{Listing}

\makeatother
\makeatletter
\@ifpackageloaded{caption}{}{\usepackage{caption}}
\@ifpackageloaded{subcaption}{}{\usepackage{subcaption}}
\makeatother

\usepackage[]{natbib}
\usepackage{bookmark}

\IfFileExists{xurl.sty}{\usepackage{xurl}}{} 
\hypersetup{
  pdftitle={Title},
  pdfauthor={Author 1; Author 2},
  pdfkeywords={3 to 6 keywords, that do not appear in the title},
  colorlinks=true,
  linkcolor={blue},
  filecolor={Maroon},
  citecolor={Blue},
  urlcolor={Blue},
  pdfcreator={LaTeX via pandoc}}

\newtheorem{theorem}{Theorem}
\newtheorem{lemma}{Lemma}

\newtheorem{proposition}{Proposition}

\def\v{{\varepsilon}}

\newcommand{\h}[1]{{\boldsymbol{#1}}}
\newcommand{\bA}{\h{A}}
\newcommand{\ba}{\h{a}}

\newcommand{\bC}{\h{C}}

\newcommand{\bI}{\h{I}}
\newcommand{\bH}{\h{H}}
\newcommand{\bM}{\h{M}}

\newcommand{\bQ}{\h{Q}}

\newcommand{\bR}{\h{R}}

\newcommand{\bT}{\h{T}}
\newcommand{\bU}{\h{U}}

\newcommand{\bX}{\h{X}}

\newcommand{\bZ}{\h{Z}}

\newcommand{\cA}{\mathcal{A}}

\newcommand{\cL}{\mathcal{L}}

\newcommand{\cN}{\mathcal{N}}
\newcommand{\cP}{\mathcal{P}}

\newcommand{\eps}{\varepsilon}

\newcommand{\bgamma}{\boldsymbol{\gamma}}

\newcommand{\bEta}{\boldsymbol{\eta}}
\newcommand{\bveps}{\boldsymbol{\eps}}

\newcommand{\bomega}{\boldsymbol{\omega}}
\newcommand{\bmu}{\boldsymbol{\mu}}
\newcommand{\bnu}{\boldsymbol{\nu}}
\newcommand{\bSigma}{\boldsymbol{\Sigma}}
\newcommand{\btau}{\boldsymbol{\tau}}
\newcommand{\brho}{\boldsymbol{\rho}}
\newcommand{\btheta}{\boldsymbol{\theta}}

\newcommand{\E}{\mathrm{E}}
\newcommand{\Var}{\mathrm{Var}}

\newcommand{\Cov}{\mathrm{Cov}}

\newcommand{\bzero}{\boldsymbol{0}}
\newcommand{\bone}{\boldsymbol{1}}

\newcommand{\anon}{1}

\begin{document}

\def\spacingset#1{\renewcommand{\baselinestretch}%
{#1}\small\normalsize} \spacingset{1}


\if1\anon
{
  \title{\bf Autocorrelation Test under Frequent Mean Shifts}
  \author[1,3]{Ziyang Liu}
\author[1,2,*]{Ning Hao}
\author[1,2]{Yue Selena Niu}
\author[4]{Han Xiao}
\author[1,3,*]{Hongxu Ding}
\affil[1]{Statistics and Data Science GIDP, University of Arizona, Tucson, Arizona, USA.}
\affil[2]{Department of Mathematics, University of Arizona, Tucson, Arizona, USA.}
\affil[3]{Department of Pharmacy Practice and Science, University of Arizona, Tucson, Arizona, USA.}
\affil[4]{Department of Statistics, Rutgers University}
\affil[*]{Corresponding Author}
\date{}                     
\setcounter{Maxaffil}{0}
\renewcommand\Affilfont{\itshape\small}
%
  \maketitle
} \fi

\if0\anon
{
  \bigskip
  \bigskip
  \bigskip
  \begin{center}
    {\LARGE\bf Title}
\end{center}
  \medskip
} \fi

\bigskip
\begin{abstract}
Testing for the presence of autocorrelation is a fundamental problem in time series analysis. Classical methods such as the Box–Pierce test rely on the assumption of stationarity, necessitating the removal of non-stationary components such as trends or shifts in the mean prior to application. However, this is not always practical, particularly when the mean structure is complex, such as being piecewise constant with frequent shifts. In this work, we propose a new inferential framework for autocorrelation in time series data under frequent mean shifts. In particular, we introduce a Shift-Immune Portmanteau (SIP) test that reliably tests for autocorrelation and is robust against mean shifts. We illustrate an application of our method to nanopore sequencing data.
\end{abstract}

\noindent%
\vfill

\newpage
\spacingset{1.8} 

\section{Introduction}

Testing for serial correlation has been a fundamental and classical problem in time series analysis. The most celebrated Box type portmanteau tests \citep{box1970distribution,ljung1978measure} and their variants \citep[see for example][for a review]{escanciano2009automatic} have become common practice for decades. This paper considers the white noise tests in the presence of frequent change points, defined as times or locations when the mean of the process shifts. Most of the existing literature on change-point analysis have been concerned with the detection of change points, see \cite{niu2016} and \cite{truong2020selective} for some overviews. However, when the temporal dependence is present, it has to be taken into account by any detection procedure \citep{tang1993effect}. While the model-based approaches \citep{davis1995testing,gombay2008change} account for it intrinsically, the problem becomes more sophisticated for nonparametric methods. \cite{altissimo2003strong,juhl2009tests,shao2010testing,aue2013structural} contain, among many others, various methods for the estimation of the variances of the test statistics or to get around it, also see \cite{perron2006dealing} for a comprehensive review of the subject.

A host of problems should and can be asked and studied in the context of change-point analysis for time series data, particularly those concerning the estimation of autocovariances \citep{tecuapetla2017autocovariance,levine2019acf} and the long run variance \citep{hall2003using,wu2007inference,khismatullina2020multiscale,chan2022optimal,bai2024difference}. In particular, \cite{tecuapetla2017autocovariance} considers the estimation of autocovariances when the mean function is piecewise smooth and the noise is $m$-dependent. Their methods rely on a high order difference of the data with difference gap $(m+1)$. In this paper, we focus on a fundamental yet less discussed one: does the serial correlation exist when the data exhibit frequent mean shifts?
We aim to address this question prior to change-point detection or autocorrelation estimation. In other words, we propose a test for the serial correlation that informs subsequent analysis based on its outcome. The validity of the proposed test is guaranteed under minimal assumptions, enhancing its utility as an initial step in data analysis.
We also highlight that the proposed test is specifically designed for data with frequent change points, such as the nanopore sequencing data that motivated our research. As demonstrated in the numerical studies in Section~\ref{sec:type1}, even a pseudo-oracle with full knowledge of the change-point locations fails to test for serial correlation if it first de-means the data segment-wise and then applies conventional white noise tests. This is not surprising: when the number of segments is large, piecewise centering introduces non-negligible artificial autocorrelations. In contrast, the proposed test effectively eliminates the impact of mean shifts, regardless of the number or magnitude of change points, ensuring a testing procedure that is faithful to its nominal level. As an additional contribution, we introduce a novel autocorrelation function (ACF) plot to visualize dependence patterns obscured by mean shifts.

This project is partially motivated by nanopore sequencing data analysis. Nanopore sequencing involves a nanoscale protein pore (nanopore) embedded in a membrane within an electrolytic solution. A consistent voltage generates an ionic current that propels negatively charged single-strand DNA or RNA sequences through the pore \citep{wang2021nanopore}. As each base passes, it disrupts the current, which is indicative of its identity. The nanopore sequencing data record these temporal currents, often consisting of tens of thousands to millions of signal points. Change-point models are often used to model the piecewise constant mean structure of nanopore sequencing data. Basecalling algorithms then translate these data into nucleotide sequences, enabling genetic analyses such as gene expression studies and mutation detection. Early basecalling methods segment raw current data and use hidden Markov models (HMMs) to identify genetic bases, while recent approaches leverage deep learning models such as recurrent neural networks (RNNs). However, these methods do not explicitly incorporate autocovariance information, despite conjectures of positive autocorrelations in nanopore sequencing data \citep{garalde2012modeling}. Inferring autocorrelation structures is challenging due to frequent mean shifts caused by the rapid molecular transit through the nanopore \citep{fleming2021nanopore}. Figure 1 illustrates examples of nanopore sequencing data with both conspicuous and subtle mean shifts, the latter being particularly difficult to detect. Developing methods to infer autocovariance structures without explicitly estimating the mean structure is critical for advancing nanopore sequencing data analysis.

\begin{figure}[h]
    \centering
    \includegraphics[width=15cm]{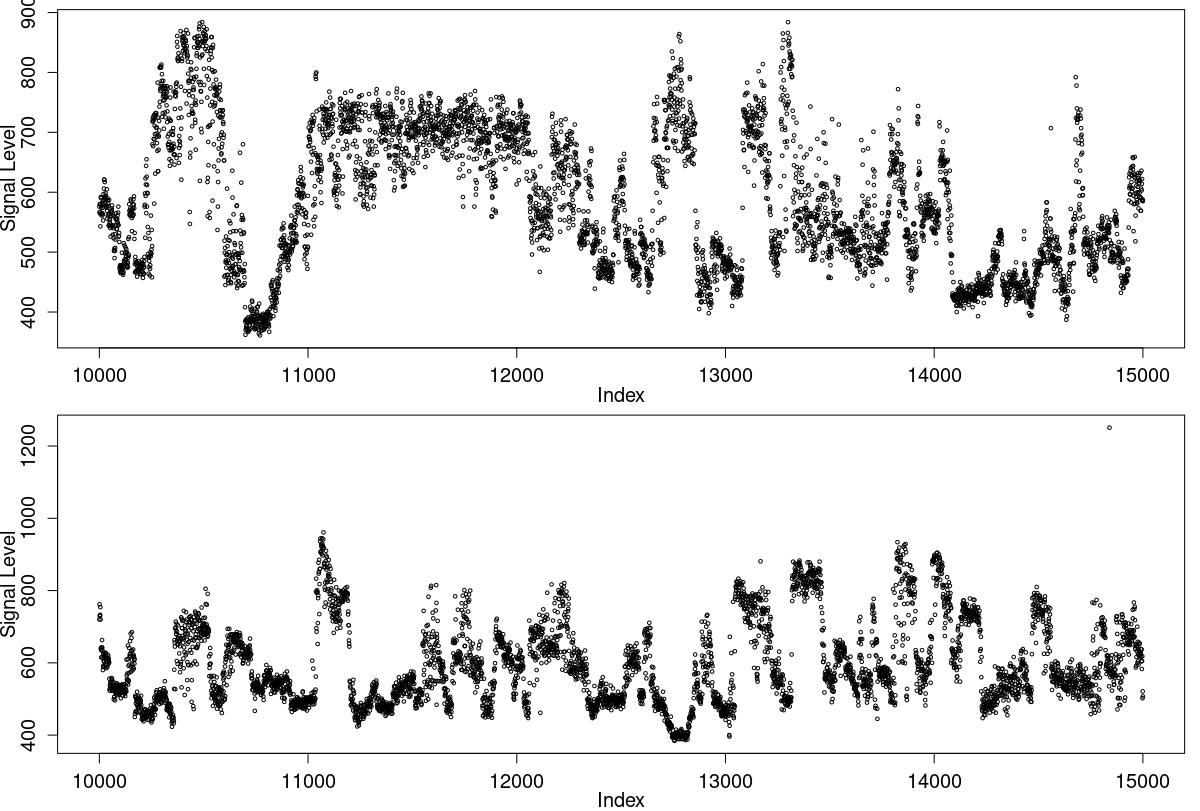}
    \caption{An illustration of nanopore sequencing data from \cite{wang2024adapting}. Top: 5000 data points from sequence id=33; bottom: 5000 data points from sequence id=39. Each of these sequences contains numerous mean shifts.}
    \label{fig1}
\end{figure}

Motivated by practical challenges and unmet methodological needs, we develop a novel statistical framework for testing and inferring covariance structures in nonstationary time series with mean shifts. Specifically, we consider all quadratic forms of the data that can be represented through symmetric Toeplitz matrices, which include the classic sample autocovariances for mean-zero time series as special cases. We then investigate whether certain quadratic statistics within this class can eliminate the impact of mean function when it is piecewise constant and the additive noise is stationary. While classical sample autocovariances become misleading in the presence of mean shifts, we identify new quadratic forms that remain robust to such shifts. Section~\ref{sec2} illustrates how the influence of mean shifts can be removed in the expected values of certain types of quadratic forms and characterizes the class of such quadratic forms. In Section~\ref{sec:SIP}, we construct a Box-type portmanteau test based on these quadratic forms and provide related asymptotic theory. In Section~\ref{sec:num}, we demonstrate the effectiveness of the proposed methods using simulated and real data examples.

In summary, this work makes several contributions to the field of time series and change-point analysis. First, it tackles the challenging problem of inferring covariance structures in nonstationary time series with frequent and irregular mean shifts. We propose novel portmanteau tests that offer robust and practical tools for analyzing nonstationary time series where traditional methods are inadequate. Our techniques would play important roles in change-point detection for time series data. Second, our visualization techniques provide an improved understanding of autocorrelation patterns in the presence of mean shifts, offering convenient tools for practitioners. Third, these contributions are particularly helpful for contemporary applications such as nanopore sequencing, where we have verified a conjecture by experts in nanopore sequencing data by showing significant positive autocorrelation in nanopore sequencing data. By addressing methodological gaps and introducing innovative tools, this work paves the way for more effective and precise analysis of nonstationary time series across a wide range of scientific disciplines.

\section{Separation of the serial correlation and mean shifts}\label{sec2}

\subsection{Model descriptions}\label{sec2.1}
Let $\{X_i\}^n_{i=1}$ be a sequence of random variables with piecewise constant means and additive stationary noises. Specifically, we assume
\begin{equation}\label{V1}
X_i =\theta_i+\v_i, \qquad  \qquad 1\leq i\leq n,
\end{equation}
where the noise sequence $\{\v_i\}^n_{i=1}$ is mean zero and stationary, and the mean parameters satisfy
\begin{equation}\label{V2}
\theta_1=\theta_2=\cdots=\theta_{\tau_1}\neq\theta_{\tau_1+1}=\cdots=\theta_{\tau_2}\neq\theta_{\tau_2+1}=\cdots\quad\cdots =\theta_{\tau_J}\neq\theta_{\tau_J+1}=\cdots=\theta_n.
\end{equation}
Let $\btau=(\tau_1,...,\tau_J)^{\top}$ be the location vector of mean shifts, which partitions the entire sequence into $J+1$ segments with constant means. For theoretical derivations, it is convenient to represent the means of these segments as $\mu_1,\dots,\,\mu_{J+1}$, respectively. Moreover, we denote the variance, lag-$h$ autocovariance and lag-$h$ autocorrelation of the noise sequence by $\gamma_0$, $\gamma_h$, and $\rho_h$, respectively.

The primary goal of this paper is to test whether $\rho_h = 0$ for all $h > 0$. In the absence of mean shifts, the process $\{X_i\}^n_{i=1}$ is stationary. The Box–Pierce test and its variants are fundamental tools for examining serial correlation. However, these methods perform poorly in the presence of mean shifts, as a non-constant mean structure can produce spurious autocorrelation, even if the underlying noises are independent. As demonstrated in our numerical studies, the type I error of the classical Box test is not well controlled when mean shifts are present.

In model \eqref{V1}, the data vector $\bX=(X_1,X_2,\ldots,X_{n})^\top$ is observed and indexed by the set of integers $[n]=\{1,...,n\}$. We write \eqref{V1} as $\bX=\btheta+\bveps$ using vector notations. For the mean vector $\btheta$, we denote by $\cL(\btheta)$ the minimal length of all constant segments in $\btheta$. That is, $\cL(\btheta)=\min_{0\leq j\leq J}\{\tau_{j+1}-\tau_j\}$ for $\btheta$ defined in \eqref{V2} with the convention $\tau_0=0$ and $\tau_{J+1}=n$.
In particular, $\cL(\btheta)=n$ implies that all observations have the same mean; $\cL(\btheta)=1$ indicates that the mean may change at two consecutive positions. For any positive integer $L\leq n$, define the set of mean vectors
\begin{equation}  \label{V3}
  \Theta_L=\{\btheta\in\mathbb{R}^n:\;\cL(\btheta)\geq L\}.
\end{equation}
For instance, $n=5$, \[\Theta_2=\{(c_1,c_1,c_1,c_1,c_1)^{\top}|c_1\in\mathbb{R}\}
\cup \{(c_2,c_2,c_3,c_3,c_3)^{\top}|c_2,\, c_3\in\mathbb{R}\}
\cup \{(c_4,c_4,c_4,c_5,c_5)^{\top}|c_4,\, c_5\in\mathbb{R}\}.\]
$\Theta_L$ with varying $L$ defines a family of parameter spaces for the mean. A higher value of $L$ in \eqref{V3} corresponds to a more restrictive model class. It is challenging to infer the autocorrelations for the model class $\Theta_L$ when $L$ is small due to frequent mean shifts.

We formalize our model description as follows.

\smallskip
\noindent {\bf Condition 1}. $\{X_i\}_{i=1}^n$ follows model \eqref{V1} with $\btheta=\E(\bX)\in\Theta_L$.

\noindent {\bf Condition 2}. The noise sequence $\{\v_i\}_{i=1}^n$ is mean-zero, strictly stationary, ergodic, and has finite fourth moments.

\noindent {\bf Condition 3}. $\max_j\{(\mu_j-\mu_{j+1})^2\}=o(n)$.

Conditions 1 and 2 restate our model assumption on the mean and variance structures. Condition 3 is mild and realistic. For example, it is reasonable to assume $\max_j\{(\mu_j-\mu_{j+1})^2\}=O(1)$ in many applications including nanopore studies. In fact, any jump of order $\log n$ or larger can be easily identified and removed; see Proposition 1 in \cite{niu2016}.

\subsection{Quadratic forms}
\label{sec:quad}

Classical Box type tests are built on the sample autocorrelations, or more fundamentally, sample autocovariances, which take the form of quadratic functions of the data. However, these standard tools are not directly applicable in our setting. To elaborate, under model \eqref{V1},
the sample autocovariance  $\frac1n\sum_{i=h+1}^n X_iX_{i-h}$  has the expectation
\begin{equation*}
    (1-h/n)\gamma_h + n^{-1}\sum_{i=h+1}^n \theta_i\theta_{i-h}.
\end{equation*}
This implies that when the data exhibit apparent autocorrelation, it is not immediately clear whether it reflects genuine temporal dependence in the noise sequence $\{\varepsilon_i\}$ or is merely an artifact of the piecewise constant mean sequence $\{\theta_i\}$. Therefore, to test for serial correlation, it is essential to isolate the contribution of the noise by filtering out the effect of the mean function.

All the sample autocovariances are quadratic forms of the data sequence, and more importantly, information about autocovariance is inherently embedded in all pairwise products $X_iX_j$. Motivated by this, we consider general quadratic forms of the type $\bX^{\top}\bA\bX$ where $\bA$ is a symmetric Toeplitz matrix of the form
\begin{align}\label{V6}
    \left(
    \begin{array}{cccccc}
         a_0 & a_1 & a_2 & \cdots & a_{n-1} \\
         a_1 & a_0 & a_1 & \cdots & a_{n-2} \\
         a_2 & a_1 & a_0 & \cdots & a_{n-3} \\
         \vdots & \vdots & \vdots & \ddots & \vdots \\
         a_{n-1} & a_{n-2} & a_{n-3} & \cdots & a_{0} \\
    \end{array}
    \right),
\end{align}
and look for suitable choices of $\bA$ as alternatives to the standard sample autocovariances, with the goal of mitigating the influence of the mean vector $\btheta$. We start with a simple observation.

\begin{proposition}\label{prop1}
Let $\bA$ be a symmetric Toeplitz matrix of the form \eqref{V6}. For $\bX$ generated from model \eqref{V1} where the noise is mean zero and stationary with a finite variance, we have
  \begin{equation}\label{V7}
      \E\left( \bX^{\top} \bA \bX\right)=\sum_{h=1-n}^{n-1} (n-|h|)a_{|h|}\gamma_h+ \btheta^{\top}\bA\btheta.
  \end{equation}
\end{proposition}

Removing the impact of the mean function amounts to ensuring that the second term $\btheta^{\top}\bA\btheta=0$ in \eqref{V7}. However, unless $\bA$ is the zero matrix, this condition cannot hold for all $\btheta\in\mathbb{R}^n$. Remarkably, the piecewise constant structure of the mean function in model \eqref{V1} allows for a class of quadratic forms whose expectations are unaffected by $\btheta$, as established by the following theorem.

\begin{theorem} \label{thm1} Let $\bA$ be a symmetric Toeplitz matrix of the form \eqref{V6}, and $L$ be an integer with $1\leq L < n/2$. We conclude that $\btheta^{\top} \bA\btheta = 0$ for all $\btheta\in\Theta_L$ if and only if the following equations hold.
\begin{align}
    a_0 + 2a_1 +\cdots + 2a_{L} =& 0,\label{V8}\\
    a_1 + 2a_2 + \cdots + La_{L} =& 0,\label{V9}\\
    a_{L+1} = a_{L+2} = \cdots = a_{n-L-1}=&0,\label{V10}\\
    La_{n-L} + (L-1)a_{n-L+1} + \cdots +a_{n-1} =& 0.\label{V11}
\end{align}
\end{theorem}

Within the class of quadratic forms characterized by Theorem~\ref{thm1}, we further consider a subclass that is invariant under a global mean shift. Specifically, we seek matrices $\bA$ such that $\bX^\top \bA \bX = (\bX + c\bone)^\top \bA (\bX + c\bone)$ for all $c \in \mathbb{R}$. This invariance property is automatically satisfied by the sample autocovariances when computed after centering the data by the sample mean. However, it requires a nontrivial calibration if the quadratic forms satisfying Theorem~\ref{thm1} are concerned.

\begin{proposition}
\label{prop:eve}
    Suppose  $L$ is an integer with $1\leq L < n/2$. Let $\bA$ be a symmetric Toeplitz matrix of the form \eqref{V6} satisfying the equations \eqref{V8}--\eqref{V11} in Theorem~\ref{thm1}. The quadratic form $\bX^\top\bA\bX$ is invariant under a global mean shift if and only if the following additional equations hold,
    \begin{equation}
        a_{i}=a_{n-i},\quad i=1,\ldots, L,
    \end{equation}
    or equivalently, $\bA$ is a circulant matrix of the form
    \begin{align}\label{eq:circA}
    \left(
    \begin{array}{cccccccccccccccc}
         a_0 & a_1 & a_2 & \cdots & a_{L-1} & a_L & 0 & \cdots & 0 & a_L & a_{L-1} & a_{L-2} & \cdots & a_2 & a_{1} \\
         a_1 & a_0 & a_1 & \cdots & a_{L-2} & a_{L-1} & a_L & \cdots & 0 & 0 & a_L & a_{L-1} & \cdots & a_3 & a_{2} \\
         a_2 & a_1 & a_0 & \cdots & a_{L-3} & a_{L-2} & a_{L-1} & \cdots & 0 & 0 & 0 & a_L & \cdots & a_4  &  a_{3} \\
         \vdots & \vdots & \vdots & \vdots & \vdots & \vdots & \vdots & \vdots & \vdots & \vdots & \vdots & \vdots & \vdots & \vdots & \vdots  \\
         a_1 & a_2 & a_3& \cdots & a_L & 0 & 0 & \cdots & a_L & a_{L-1} & a_{L-2} & a_{L-3} & \cdots & a_1 & a_{0} \\
    \end{array}
    \right).
    \end{align}
\end{proposition}

To sum up, in order for the quadratic form $\bX^\top\bA\bX$ to be both independent of $\btheta$ in expectation when $\btheta\in\Theta_L$ and invariant under a global mean shift, $\bA$ must be a circulant matrix with the first row $(a_0,a_1,\ldots,a_L, 0, \cdots, 0, a_L,\cdots,a_2,a_1)$, where the coefficients $a_0,a_1,\ldots,a_L$ satisfy \eqref{V8} and \eqref{V9}. We denote the set of all such $\bA$ matrices by $\mathcal A_L$.
When $\bA\in\cA_L$, the quadratic form $\bX^\top\bA\bX$ has a more explicit representation.
\begin{proposition}\label{prop:T rep}
For $\bA\in\cA_L$ of the form in \eqref{eq:circA}, we have
  \begin{equation}
    \label{eq:XAX_T}
    \bX^\top\bA\bX = -\sum_{h=1}^L a_hT_h,
\end{equation}
where $T_h=\sum_{i=1}^{n}(X_i-X_{i+h})^2$,  $h=1,\dots, L.$
\end{proposition}

Our approach is to base the test for serial correlation on the quadratic forms, so it is also necessary to remove the variance $\gamma_0$ from \eqref{V7}.

\begin{proposition}
\label{prop:XAX_autocov}
    Assume Conditions 1 and 2. Suppose $\bA\in\cA_L$, the expectation of $\bX^\top\bA\bX$ does not depend on $\gamma_0$ if and only if $a_0=0$, which is equivalent to $a_1+\cdots+a_L=0$.
\end{proposition}

Therefore, we have narrowed the choice of the quadratic forms down to the class $\cA^\circ_L$, represented by vectors in $\mathbb{R}^L$ as follows.
\begin{equation}
    \label{eq:A_0}
    \cA^\circ_L = \{(a_1,\ldots,a_L)^\top\in\mathbb{R}^L \mid a_1+\cdots+a_L=0,\;\;a_1+2a_2+\cdots+La_L=0\}.
\end{equation}
By this convention, there is a natural embedding $\cA^\circ_K\subset \cA^\circ_L$ when $K<L$, which maps $(a_1,\ldots,a_K)^{\top}$ to  $(a_1,\ldots,a_K,0,\ldots,0)^{\top}$. In other words, $\{\cA^\circ_L\}$ represents a nested family of quadratic forms.
For any $\ba=(a_1,\ldots,a_L)^\top\in\cA^\circ_{L}$, we use $\bA_{\ba}$ to denote the corresponding circulant matrix starting from the row $(0,a_1,\ldots,a_L, 0, \cdots, 0, a_L,\cdots,a_2,a_1)$, and $\bX^{\top}\bA_{\ba}\bX$ to denote the corresponding quadratic forms.

\section{Shift-Immune Portmanteau Test }\label{sec:SIP}

\subsection{Construction of the Test Statistic}
\label{sec:TS}

In designing the test, we are primarily motivated by data that may exhibit frequent changes in the mean function $\btheta$. The class of quadratic forms discussed in the previous section captures information about the autocovariances while eliminating the influence of $\btheta$, thus serving as the building blocks of the proposed test statistic. Before proceeding further, we remark that the discussion in Section~\ref{sec:quad} relies on the key quantity $L$, the minimal segment length of $\btheta$. Our strategy is to construct the test statistic based on the class $\cA^\circ_{m+2}\subset\cA^\circ_L$ with the assumption $m+2\leq L$ or $m+2\leq L/2$, where $m$ is a pre-specified number of lags, as in all Box-type tests. Since traditional Box-type tests typically involve a modest number of autocorrelations, we do not aim to use a large $m$ either. We believe that $m+2\leq L$ or $m+2\leq L/2$ is a reasonable assumption and do not require any additional conditions  on $L$.

With a specified number of lags $m$, we focus on the quadratic forms parametrized by the linear space $\cA^\circ_{m+2}$. As shown in \eqref{V7}, if $\btheta\in\Theta_L$, then for any $\ba\in\cA^\circ_{m+2}\subset\cA^\circ_L$,
\begin{equation*}
    \E \left(\bX^\top\bA_{\ba}\bX\right) = \sum_{h=-(m+2)}^{m+2}(n-|h|)a_{|h|}\gamma_h + \sum_{h=-(m+2)}^{m+2} |h|a_{|h|}\gamma_{n-|h|}.
\end{equation*}
The expectation involves two sets of autocovariances $\{\gamma_1,\ldots,\gamma_{m+2}\}$ and $\{\gamma_{n-1},\gamma_{n-2},\ldots,\gamma_{n-m-2}\}$. Under any short-range dependence condition, the second set becomes negligible when the sample size grows, so we will essentially use the quadratic form to test whether the first set of autocovariances are zero or not.

It is evident that $\mathcal A^\circ_{m+2}$ is a $m$-dimensional linear subspace of $\mathbb{R}^{m+2}$. To accumulate the information on $\{\gamma_1,\ldots,\gamma_{m+2}\}$, we pick a set of linearly independent elements $\{\ba_h,\;h=1,2,\ldots,m\}$ from $\cA^\circ_{m+2}$, and consider the corresponding quadratic forms $\{\bX^\top\bA_{\ba_h}\bX\}$. Define an $(m+2)\times m$ matrix $\bC=[\ba_1,\ba_2,\ldots,\ba_m]$ and an $(m+2)$-dimensional vector $\bT_{m+2}=(T_1,\ldots,T_{m+2})^\top$. In view of \eqref{eq:XAX_T} in Proposition \ref{prop:T rep}, these quadratic forms constitute the random vector $\bC^\top\bT_{m+2}$. Since the expectation of this random vector involves linear combinations of $\{\gamma_1,\ldots,\gamma_{m+2}\}$, should they be nonzero, it is natural to ``square'' $\bC^\top\bT_{m+2}$ to form the test statistic. As will be elaborated in Lemma \ref{lemma1}, the random vector $\bT_{m+2}$ has nonzero covariances induced by the mean structure $\btheta$, even under the null hypothesis that the $\varepsilon_i$'s are independent and identically distributed (IID). Therefore, the random vector $\bC^\top\bT_{m+2}$  must be properly scaled with respect to its covariance matrix when forming the test statistic.

Let $\bSigma$ be any nonsingular $(m+2) \times (m+2)$ covariance matrix. If $\bSigma$ is the covariance matrix of $\bT_{m+2}$, then the covariance matrix of $\bC^\top \bT_{m+2}$ is given by $\bC^\top \bSigma \bC$. The test statistic that we will propose takes the form:
\begin{equation}
    \label{eq:TS}
\bT_{m+2}^\top\h{C}\left[\h{C}^\top\bSigma\h{C}\right]^{-1}\h{C}^\top\bT_{m+2}.
\end{equation}

A more detailed discussion on the calculation and estimation of the test statistic is deferred to Section~\ref{sec:asymp}. Here, we simply note that the test statistic \eqref{eq:TS} does not depend on any specific choice of the basis ${\ba_h}$ of $\mathcal{A}^\circ_{m+2}$ or the matrix $\bC$.
Suppose $\bU$ is a $(m+2)\times m$ matrix whose columns form another basis of $\cA^\circ_{m+2}$, then we can always write $\bC$ as $\bU\bQ$ where $\bQ$ is a non-singular square matrix. Replacing $\bC$ by $\bU\bQ$, the preceding equation becomes
\begin{equation}
\label{eq:TS_invariance}
    \bT_{m+2}^\top\bU\bQ \left[\bQ^\top\h{U}^\top\bSigma\h{U}\bQ\right]^{-1}\bQ^\top\h{U}^\top\bT_{m+2} = \bT_{m+2}^\top\h{U}\left[\h{U}^\top\bSigma\h{U}\right]^{-1}\h{U}^\top\bT_{m+2}.
\end{equation}
Therefore, the quantity in \eqref{eq:TS} does not depend on any specific choice of $\bC$ as long as its column space spans $\cA^\circ_{m+2}$. In other words, our proposed test statistic is unique for each $m$.

\subsection{Choice of Basis}
\label{sec:basis}

The aforementioned invariance property with respect to $\bC$ allows us to choose any suitable $\bC$ to facilitate the subsequent discussion. For the sake of interpretability,  we make the following choice of $\ba_h=(a_{h1},a_{h2},\ldots,a_{h,m+2})^\top$, $h=1,2,\ldots,m$:
\begin{equation}
    \label{toep2}
\begin{aligned}
 & a_{hh}=\frac{1}{2n},\; && a_{h,m+1}=-\frac{m+2-h}{2n}\\
 & a_{h,m+2}=\frac{m+1-h}{2n},\; && a_{hi}=0\, \text{ for } i\notin\{h,\,m+1,\,m+2\}.
\end{aligned}
\end{equation}

Such a choice can be motivated by presuming an additional $m$-dependence structure of the noise sequence $\{\varepsilon_i\}$ on top of Condition 2. We emphasize that this $m$-dependence assumption is introduced solely for the interpretation of the construction in \eqref{toep2}; it is not required for the validity of our test procedure. For each $1\leq h\leq m$, let $\bA_h$ be the circulant matrix induced by $\ba_h$. We begin by explicitly expanding the quadratic form $\bX^\top \bA_h \bX$ and assigning it the name
\begin{equation}\label{gamma_hat1}
\begin{aligned}
        \hat\gamma_h & :=\bX^\top\bA_h\bX \\
        & = \frac{1}{n}\left(\sum^{n}_{i=1}X_i X_{i+h} -(m+2-h)\sum^{n}_{i=1}X_i X_{i+m+1} + (m+1-h)\sum^{n}_{i=1}X_i X_{i+m+2}\right) \\
        & = (2n)^{-1}\left[-T_h + (m+2-h )T_{m+1} - (m+1-h)T_{m+2}\right],
\end{aligned}
\end{equation}
where we use the convention that $X_{n+i}=X_i$ whenever the subscript goes beyond $n$.
We further define:
\begin{equation}
    \hat{\gamma}_0 = (2n)^{-1}\left[(m+2)T_{m+1} -(m+1)T_{m+2}\right]. \label{gamma0 hat}
\end{equation}

\begin{proposition} \label{Prop3}
Under Conditions 1 and 2, if $m+2\leq L$, $m+2<n/2$, and that $\{\varepsilon_i\}$ is $m$-dependent, then
    \begin{align*}
        \E \hat\gamma_0 & =\gamma_0, \\
        \E \hat\gamma_h & = (1-h/n)\gamma_h,\; && 1\leq h\leq m.
    \end{align*}
\end{proposition}
The proposition confirms that $\{\hat\gamma_h\}$ are valid estimates of the corresponding autocovariances under the $m$-dependence condition, thus justifying their appellations. More importantly, they are all valid under the presence of any mean function $\btheta\in\Theta_L$, and are invariant under the global mean shift. Consequently, we define
\begin{equation}
    \label{eq:rhohat}
    \hat\rho_h=\hat\gamma_h/\hat\gamma_0, \quad h\geq 1.
\end{equation}

\subsection{What is being tested?}
In this subsection. we investigate when the test based on \eqref{eq:TS} and the equivalent form \eqref{eq:TS_invariance} has power asymptotically. In view of \eqref{eq:A_0}, $\cA^\circ_{m+2}$ is naturally a linear subspace of $\mathbb{R}^{m+2}$. Let $\cP_{m+2}$ denote the orthogonal projection matrix onto the subspace $\cA^\circ_{m+2}$.
\begin{proposition}
    \label{prop:PT}
    Under Conditions 1, 2, and 3, if $m$ an integer such that $m+2\leq L$, and $m+2<n/2$, then
    \begin{equation}
        \label{eq:PT}
        \cP_{m+2}\bT_{m+2}/(2n)\xrightarrow{P}\cP_{m+2}\bgamma_{m+2},
    \end{equation}
    where $\bgamma_{m+2}=(\gamma_1,\gamma_2,\ldots,\gamma_{m+2})^\top$.
\end{proposition}

Consequently, the test statistic in \eqref{eq:TS} essentially tests whether the projection $\cP_{m+2} \bgamma_{m+2}$ is zero. Indeed, \eqref{eq:TS} can be rewritten as
\begin{equation*}
\bT_{m+2}^\top\bC\left[\bC^\top\bSigma\bC\right]^{-1}\bC^\top\bT_{m+2}=\left(\cP_{m+2}\bT_{m+2}\right)^\top\bC\left[\bC^\top\bSigma\bC\right]^{-1}\bC^\top\left(\cP_{m+2}\bT_{m+2}\right),
\end{equation*}
since $\cP_{m+2}\bC=\bC$. This implies that the test would have no power if $\cP_{m+2}\bgamma_{m+2}=\bzero$. This is the price we pay for eliminating the influence of the nontrivial piecewise-constant mean vector $\btheta$, and it merits further discussion. In classical time series analysis, especially in diagnostics, when checking whether a series has autocorrelations, it has been customary to apply Box-type tests for multiple values of $m$, as has been rendered in some standard {\tt R} packages, e.g. the {\tt tsdiag()} function in the base {\tt stats} package. If $\cP_{m+2}\bgamma_{m+2}=\bzero$ for every $m$, then it entails that, in view of \eqref{toep2}, $\gamma_m-2\gamma_{m+1}+\gamma_{m+2}=0$ for every $m\geq 1$, or equivalently,
\[\gamma_{m+2}-\gamma_{m+1}=\gamma_{m+1}-\gamma_m,\text{ for all }m\geq 1.\] The general solution of this difference equation takes the form $\gamma_h=c_0+c_1h$, where $c_0$ and $c_1$ are constants. However, such a linear form for the autocovariance sequence ${\gamma_h}$ is incompatible with the short-range dependence condition $\gamma_h \to 0$ as $h \to \infty$. Therefore, under this condition, the test must have asymptotic power for some $m$, provided that not all $\gamma_h$ are zero. We thus conclude that the loss of power due to removing the impact of $\btheta$ is not substantial.

\subsection{Asymptotic theory}
\label{sec:asymp}

Before stating the asymptotic result, we introduce some notations. Let $\kappa_4=\E(\v_1^4)/\gamma_0^2$. Define $w=W(\btheta)/(n\gamma_0)$, where
\begin{equation*}
W(\btheta) = \sum_{i=1}^{n}(\theta_i-\theta_{i+1})^2=\sum_{j=1}^{J+1}(\mu_j-\mu_{j+1})^2.
\end{equation*}
Let $\bI_m$ be the $m\times m$ identity matrix, $\bH_m=(H_{ij})$ be an $m\times m$ matrix with $H_{ij}=\min\{i,j\}$,  $\boldsymbol{\eta}_m$ be the vector $(1,2,\ldots,m)^\top$, $\bzero_m$ and $\bone_m$ be vectors of length $m$ with all entries equal to 0 and 1, respectively.
In the asymptotic analysis, we treat $\btheta$ and its characteristics such as $J$, $\bmu$, and $w$ as functions of $n$. 

As discussed in Section~\ref{sec:TS}, the test statistic \eqref{eq:TS} does not depend on the particular choice of $\bC$, as long as its column space is the same as $\cA_{m+2}^\circ$. Therefore, we adopt the version of $\bC$ described in \eqref{toep2} in Section~\ref{sec:basis}, and use the notations $\hat\gamma_h$ and $\hat\rho_h$ defined in \eqref{gamma_hat1} and \eqref{eq:rhohat}, which facilitate the formulation of the results.  We begin with a central limit theorem for $\hat\bgamma_{m}=(\hat\gamma_1,\ldots,\hat\gamma_m)^\top$. Since the variance of $\hat\bgamma_m$ depends on $w$, which is allowed to vary as $n$ grows, we need to normalize $\hat\bgamma_m$ by its asymptotic covariance matrix (scaled by $1/n$)
\begin{align}\label{T1}
  \bSigma_{\gamma,w} : =&  \gamma_0^2\{\bI_m+\left[(2m^2+6m+5)+2(m^2+3m+2)w\right]\bone_m\bone_m^{\top}\nonumber\\
   \quad & -\left[(2m+3)+2(m+2)w\right](\boldsymbol{\eta}_{m}\bone_{m}^{\top} +\bone_{m}\boldsymbol{\eta}^{\top}_{m})\nonumber\\&+(2+2w)\boldsymbol{\eta}_{m}\boldsymbol{\eta}^{\top}_{m} +2w\bH_m\}.
\end{align}
We append the subscript $w$ to $\bSigma_{\gamma,w}$ to emphasize its dependence on $w$.

\begin{theorem}\label{thm2}
Under conditions 1, 2, and 3,  and the null hypothesis that $\eps_1, \dots, \, \eps_n$ are IID, suppose $m$ is an integer such that $m+2\leq L/2$, and $m+2<n/2$, then $\hat\gamma_0$ and $\hat{\bgamma}$ satisfy
\begin{equation*}
    \hat{\gamma}_0 \xrightarrow{P}\gamma_0, \quad
    \bSigma_{\gamma,w}^{-1/2}(\sqrt{n}\hat{\bgamma}_m)\xrightarrow{D}  \cN(\bzero_{m}, \bI_{m}),\quad \text{as }   n\to\infty.
\end{equation*}
Moreover, under a relaxed condition $m+2\leq L$, it holds that
\begin{equation*}
    \lim_{n\rightarrow\infty}\bSigma_{\gamma,2w}^{-1/2}(\sqrt{n}\hat{\bgamma}_m) \preceq  \cN(\bzero_{m}, \bI_{m}).
\end{equation*}
\end{theorem}

The notation $\preceq$ in the second statement of Theorem~\ref{thm2} indicates stochastic dominance. Specifically, for a sequence of random vectors $\{\h{Q}_n\}$ and a ``limit'' random vector $\h{Q}$, the statement $\lim_{n\rightarrow\infty}\h{Q}_n\preceq \h{Q}$  means that
\begin{equation*}
    \mathop{\overline{\lim}}_{n\rightarrow\infty} \, P\left[\left|\h{b}^\top\h{Q_n}\right|\geq x\right] \leq P\left[\left|\h{b}^\top\h{Q}\right|\geq x\right],
\end{equation*}
for any unit vector $\h{b}$ and any positive number $x$.

Let $\hat\brho_m=(\hat\rho_1,\ldots,\hat\rho_m)^\top=\hat\bgamma_m/\hat\gamma_0$. It follows Theorem \ref{thm2} and Slutsky's theorem that  $\bSigma_{\rho,w}^{-1/2}(\sqrt{n}\hat{\brho}) \to \cN(\bzero_{m}, \bI_m)$, where
$\bSigma_{\rho,w}=\bSigma_{\gamma,w}/\gamma_0^2$, and $\bSigma_{\gamma,w}$ is defined in \eqref{T1}. In fact, this asymptotic property holds for any autocorrelation estimators
$\hat\bgamma/\tilde{\gamma}_0$ whenever $\tilde{\gamma}_0$ is consistent. Moreover, the asymptotic covariance $\bSigma_{\rho,w}$ depends on only one unknown quantity $w$, which makes it easy to estimate. For example, we could employ
\begin{equation}\label{w1}
\hat w_1 = 2(n\hat\gamma_0)^{-1}\left(\sum^{n}_{i=1}X_i X_{i+m+2}-\sum^{n}_{i=1}X_i X_{i+m+1}\right) = (n\hat\gamma_0)^{-1} (T_{m+2}-T_{m+1}).
\end{equation}

In addition, under the null hypothesis, $w$ can be estimated from a regression model studied in \cite{hao2023equivariant}
\begin{equation*}\label{Th}
  (2n)^{-1}{T_h}=\alpha+h\beta+e_h, \qquad h=1, \dots, m+2,
\end{equation*}
where $(\alpha, \beta)^{\top}=(\gamma_0, w\gamma_0/2)^{\top}$.
The least squares estimator is defined as
\begin{equation}\label{ols}
(\hat{\alpha},\hat\beta)^{\top} =(\bZ_{m+2}^{\top}\bZ_{m+2})^{-1}\bZ_{m+2}^{\top}\bT_{m+2}/2n,
\end{equation}
where $\bZ_{m+2}=(\bone_{m+2}, \bEta_{m+2})$. Consequently, $w$ can be estimated by
\begin{equation*}\label{w hat}
    \hat{w}_2 =2\hat{\beta}/\hat{\alpha}.
\end{equation*}

The empirical version of $\bSigma_{\rho,w}$ is defined by plugging $\hat w$ ($\hat w_1$ or $\hat w_2$) in $\bSigma_{\rho,w}=\bSigma_{\gamma,w}/\gamma_0^2$, i.e.,
\begin{align}\label{hat_sigma_rho}
  {\bSigma}_{\rho,\hat w}=&  \bI_m+\left[(2m^2+6m+5)+2(m^2+3m+2)\hat{w}\right]\bone_m\bone_m^{\top}\notag\\
   \quad & -\left[(2m+3)+2(m+2)\hat{w}\right](\boldsymbol{\eta}_{m}\bone_{m}^{\top} +\bone_{m}\boldsymbol{\eta}^{\top}_{m})\notag\\&+(2+2\hat{w})\boldsymbol{\eta}_{m}\boldsymbol{\eta}^{\top}_{m} +2\hat{w}\bH_m.
\end{align}

\begin{theorem}\label{thm3}
Under conditions 1, 2, and 3,  and the null hypothesis that $\eps_1, \dots, \, \eps_n$ are IID, if $m+2\leq L/2$, and $m+2<n/2$, then
\begin{equation} \label{chi2}
 n\hat{\brho}^{\top}{\bSigma}_{\rho,\hat w}^{-1}\hat{\brho} \xrightarrow{D} \chi^2_m,
\end{equation}
where ${\bSigma}_{\rho,\hat w}$ is defined in \eqref{hat_sigma_rho}, $\hat\brho=\hat\bgamma/\hat\gamma_0$ for any consistent estimator $\hat\gamma_0$.

Moreover, under a relaxed condition $m+2\leq L$, it holds that
\begin{equation*}
 \lim_{n\rightarrow\infty}n\hat{\brho}^{\top}{\bSigma}_{\rho,2\hat w}^{-1}\hat{\brho} \preceq \chi^2_m.
\end{equation*}
\end{theorem}

\subsection{Test procedures}
\label{sec:test}

In light of Theorem \ref{thm3}, we propose using $n\hat{\brho}_m^{\top} {\bSigma}_{\rho,\hat w}^{-1}\hat{\brho}_m$ as a test statistic. At this point it is clear that the proposed test statistic bears much resemblance with the classical Box-type portmanteau test \citep{box1970distribution}, with the important difference that the quadratic norm of $\hat\brho_m$ has to be calculated under a proper precision matrix ${\bSigma}_{\rho,\hat w}^{-1}$. The construction of the test statistic, including the calculation of $\hat\rho_m$ and its covariance matrix, is motivated and necessitated by data with frequent mean shifts. We therefore refer to $n\hat{\brho}_m^{\top} {\bSigma}_{\rho,\hat w}^{-1}\hat{\brho}_m$ as the Shift-Immune Portmanteau (SIP) test statistic. Based on the theoretical results, we introduce two versions of the SIP test below.

Similar to the Box tests, we need to preset the order of autocorrelation $m$ to construct the test statistic. Given $m$, we estimate the vector of $\hat\bgamma_m$ via \eqref{gamma_hat1}. The statistic $\hat\brho$ then depends on the choice of a consistent estimator for $\gamma_0$. In addition, the test statistic relies on an estimator to $w$. We consider two estimators for the couple $(\gamma_0,w)$. First, we can use the estimator $\hat\gamma_0$ in \eqref{gamma0 hat} and $\hat w_1$ in \eqref{w1}. Alternatively, we can use the EVE estimator $\hat\alpha$ \eqref{ols} for $\gamma_0$ and the associated $\hat w_2$ for $w$. These two options lead to two test statistics, each of which follows a $\chi^2_m$ distribution asymptotically under the null hypothesis, as guaranteed by Theorem \ref{thm3}. We refer to these two test statistics as SIP 1 and SIP 2, respectively. As our numerical studies will show, both methods perform well overall. However, SIP 2 provides more accurate control of the type I error rate when $m$ is large.

\subsection{Shift-Immune ACF plot}

The Autocorrelation Function (ACF) plot is a valuable tool in time series analysis that visualizes correlations between observations at successive time lags. In \texttt{R}, the ACF plot is typically generated using the \texttt{acf()} function, which displays the strength of autocorrelation at various lags together with 95\% significance bounds (under the IID assumption). By examining the ACF plot, practitioners can evaluate stationarity (based on how quickly autocorrelations decay), identify potential seasonal patterns through recurring spikes, and determine significant lags for stationary series. It is worth noting that the ACF plot provides straightforward, reliable information on autocorrelations only if the time series is stationary; otherwise, it serves primarily as a diagnostic tool.

Our framework provides a useful tool to visualize autocorrelation for nonstationary time series with frequent mean shifts. The main ingredients of an ACF plot include an estimated autocorrelation and a significance bound for each lag of interest. Due to the non-constant nature of $\btheta$, it is impossible to estimate the autocovariances consistently without resorting to assuming a further dependence structure (like the $m$-dependence in \cite{tecuapetla2017autocovariance}) or to assuming $L$ goes to infinity to allow the asymptotic analysis. As emphasized in the introduction, we intend the exposition in this paper to serve as an initial step of the analysis, and be valid under as mild conditions as possible. Therefore, we do not try to estimate the ACF to generate the plot. Instead, for a prefixed maximal lag $s$, we use \eqref{gamma_hat1} and \eqref{gamma0 hat} with $m=h+2$ to calculate $\hat\gamma_0$, $\hat\gamma_h$, and $\tilde\rho_h=\hat\gamma_h/\hat\gamma_0$ for each $h=1,\dots,\, s$. The significance bounds are derived from the diagonal elements in $\bSigma_{\rho,\hat w}$  in \eqref{hat_sigma_rho}, with $\hat w=\hat w_1$, as defined in \eqref{w1} with $m=s$. The advantage of this approach is that the standard error of $\hat\rho_h$ under IID assumption (or more general, $h$-dependent assumption on $\{\varepsilon_i\}$) is $\sqrt{(6+4\hat w)/n}$, independent of $h$, as verified by the following proposition.

\begin{proposition}\label{prop5}
  Assume the conditions of Theorem~\ref{thm3}. The $(m,m)$ entry of the $m\times m$ matrix $\bSigma_{\rho,\hat w}$ defined in \eqref{hat_sigma_rho} is $6+4\hat w$.
\end{proposition}

We will then plot $\tilde\rho_h$ against $h$ for $h=1,2,\ldots,s$ with the corresponding 95\% significance bounds $\pm1.96\sqrt{(6+4\hat w)/n}$. We remark that the $\hat\gamma_h$ generated this way is actually estimating $\gamma_h-2\gamma_{h+1}+\gamma_{h+2}$ (see Proposition~\ref{prop:PT}), which is unbiased only when $\gamma_{h+1}=\gamma_{h+2}=0$. Therefore, the resulting plot is not an ACF plot in the strict sense. Nevertheless, it provides valuable information about autocorrelation patterns in the presence of mean shifts, and so we adopt the term ``ACF plot'' by a slight abuse of terminology. When the plot reveals more than one significant autocorrelation, the estimated autocorrelation at the highest lag likely gives a faithful estimate.

\section{Numerical Studies}
\label{sec:num}

\subsection{Simulated data examples}

\label{sec:sim}

\subsubsection{Type I error control}
\label{sec:type1}

We illustrate the performance of our proposed test procedures using simulated data.
We consider three test procedures, including our proposed methods SIP 1, SIP 2, and the classical Box-Pierce test (labeled as `Box'). We also add two oracle procedures for comparison. Specifically, assuming the true mean is known, the `oracle' directly applies the Box-Pierce test to the noise sequence. In the pseudo-oracle procedure (labeled as `p-oracle'), assuming the locations of the change points are known, the Box-Pierce test is applied to the residuals, which are the differences between observed values and the segment-wise sample means. For all test procedures, the order $m$ (the \texttt{lag} parameter of the \texttt{Box.test()} function in \texttt{R}) is chosen from the set $\{1,2,4,8\}$.

We consider three scenarios for the noise distribution: a standard Gaussian distribution $\v_i\sim N(0,1)$, a scaled student's $t$-distribution with 6 degrees of freedom $\v_i\sim \sqrt{\tfrac{2}{3}} \,t_6$, and a translated exponential distribution $\v_i\sim Exp(1)-1$. All distributions have a mean of zero and variance of one. For the mean structure, we fix $n = 10,000$, $J = 100$, and $\btheta\in\Theta_{20}$. We randomly generate 100 change-point locations, ensuring that $\cL(\btheta) \geq 20$. Additionally, the segment mean parameters are drawn independently from a uniform distribution over the interval $[-5, 5]$. This mean vector is generated and used in all scenarios.  In Table \ref{type I error}, we list the average empirical type I error rates over 1,000 independent replicates.

\begin{table}[ht]
\begin{center}
\caption{Estimated Type I error rates
with 1000 replicates under three noise distributions.}\label{type I error}
\begin{tabular}{llcccc}
\hline
& &$m=1$&$m=2$&$m=4$&$m=8$\\
\hline
Gaussian &SIP 1& 0.047 &0.034&0.050  &0.080  \\
&SIP 2&0.050& 0.031& 0.044& 0.048\\
&Box&1.000&1.000&1.000&1.000\\
&oracle&0.050&0.057&0.044&0.048\\
&p-oracle&  0.172& 0.233& 0.322& 0.412\\
\hline
exponential &SIP 1&0.052 &0.042  &0.060  &0.103  \\
& SIP 2&0.044& 0.040 &0.049 &0.048\\
&Box&1.000&1.000&1.000&1.000\\
&oracle&0.045&0.058&0.052&0.054\\
&p-oracle& 0.155& 0.201 &0.274 &0.421\\
\hline

student's $t$ &SIP 1&0.046 &0.045  &0.053  &0.085  \\
&SIP 2&0.043 &0.045& 0.051& 0.059\\
&Box&1.000&1.000&1.000&1.000\\
&oracle&0.048&0.046&0.047&0.043\\
&p-oracle& 0.174& 0.215& 0.260 &0.407\\
\hline
\end{tabular}
\end{center}
\end{table}

We observe that the Box-Pierce test consistently yields a type I error rate of 1, indicating the significant impact of frequent mean shifts on the test for serial correlation. In contrast, the SIP methods demonstrate excellent control of the type I error across different noise distributions, suggesting that their performance is robust to different noise characteristics, including heavy-tailed or asymmetric distributions. This behavior is consistent with the Box-Pierce test for stationary time series. The SIP 1 test  controls the type I error well for all considered values of $m$, except when $m=8$. This is primarily due to the instability of the variance estimator $\hat{\gamma}_0$ for larger values of $m$, which compromises the type I error control. On the other hand, the SIP 2 test performs as effectively as the oracle procedure across all $m$ values.

A noteworthy finding is that even when the true locations of the change points are known, the pseudo-oracle procedure fails to control the type I error. This failure arises because the demeaning step introduces significant autocorrelation under the scenario with frequent mean shifts. This suggests that, even when change points can be accurately detected, the SIP method remains the preferred choice for testing serial correlation.

\subsubsection{Power Analysis}
We illustrate the statistical power of the proposed SIP tests and compare them with the oracle procedure. We use the same mean vector $\btheta$ as in the type I error analysis, and consider three noise structures detailed below, with $\{z_i\}$ drawn IID from the standard normal distribution.

Noise structure 1: a first-order moving-average model (MA(1)) where $\eps_i = z_i + \omega z_{i-1}$.

Noise structure 2: a fourth-order moving-average model (MA(4)) where $\eps_i = z_i + \sum^4_{j=1}\omega_j z_{i-j}$.

Noise structure 3: a first-order auto-regressive model (AR(1)) where $\eps_i = \phi\eps_{i-1} +z_i$.

For the MA(1) case, we vary $\omega$ in $\{-0.1, -0.05, -0.025, 0.025, 0.05, 0.1\}$ to examine the impact of the sign and magnitude of autocorrelation. For the MA(4) case, we consider three scenarios for $\bomega=(\omega_1,\dots,\omega_4)$: $(0.5, 0.4, 0.3, 0.2)$, $(0.1, 0.1, 0.5, -0.4)$, and $(0, 0.1, 0, -0.8)$, with corresponding autocorrelations $\brho$ of $(0.571, 0.409, 0.260, 0.130)$, $(0.028, 0.077, 0.322, 0.284)$, and $(0, 0.012, 0, -0.485)$. For the AR(1) case, we select $\phi$ from $\{-0.1, -0.05, -0.025, 0.025, 0.05, 0.1\}$. We report the average power of the SIP methods and the oracle procedure using 1,000 replicates in Tables \ref{power MA1}-\ref{power AR1}.

\begin{table}[ht]
\begin{center}
\caption{Average power under MA(1) noise with 1,000 replicates and varying $\omega$.}\label{power MA1}
\begin{tabular}{llcccc}
\hline
& method &$m=1$&$m=2$&$m=4$&$m=8$\\
\hline
$\omega=-0.1$& SIP 1&0.929&0.988&0.999&0.998\\
 & SIP 2&0.919&0.987&0.999&0.999\\
 & oracle&1.000&1.000&1.000&1.000\\
\hline
$\omega=-0.05$& SIP 1&0.469&0.598&0.675&0.644\\
 & SIP 2&0.433&0.578&0.652&0.655\\
 & oracle&0.996&0.993&0.986&0.966\\
\hline
$\omega=-0.025$& SIP 1&0.161&0.203&0.224&0.216\\
 & SIP 2&0.142&0.176&0.196&0.184\\
 & oracle&0.720&0.599&0.474&0.352\\
\hline
$\omega=0.025$& SIP 1&0.145&0.161&0.165&0.161\\
 & SIP 2&0.172&0.178&0.180&0.148\\
 & oracle&0.671&0.567&0.452&0.350\\
\hline
$\omega=0.05$& SIP 1&0.491&0.606&0.671&0.635\\
 & SIP 2&0.531&0.632&0.706&0.687\\
& oracle&0.998&0.998&0.987&0.964\\
\hline
$\omega=0.1$& SIP 1&0.984&0.998&1.000&0.999\\
& SIP 2&0.987&0.998&1.000&1.000\\
& oracle&1.000&1.000&1.000&1.000\\
\hline
\end{tabular}
\end{center}
\end{table}

\begin{table}[ht]
\begin{center}
\caption{Average power under MA(4) noise with 1,000 replicates and three scenarios: Scenario 1, $\bomega= (0.5, 0.4, 0.3, 0.2)$; Scenario 2, $\bomega=(0.1, 0.1, 0.5, -0.4)$, and Scenario 3, $\bomega=(0, 0.1, 0, -0.8)$.}\label{power MA4}
\begin{tabular}{llcccc}
\hline
&method&$m=1$&$m=2$&$m=4$&$m=8$\\
\hline
Scenario 1& SIP 1&0.249&0.874&1.000&1.000\\
 & SIP 2&0.282&0.918&1.000&1.000\\
 & oracle&1.000&1.000&1.000&1.000\\
\hline
Scenario 2& SIP 1&1.000&1.000&1.000&1.000\\
 & SIP 2&1.000&1.000&1.000&1.000\\
 & oracle&0.741&1.000&1.000&1.000\\
\hline
Scenario 3& SIP 1&0.104&1.000&1.000&1.000\\
 & SIP 2&0.078&1.000&1.000&1.000\\
 & oracle&0.097&0.143&1.000&1.000\\
\hline
\end{tabular}
\end{center}
\end{table}

\begin{table}[ht]
\begin{center}
\caption{Average power under AR(1) noise with 1,000 replicates and varuing $\phi$.}\label{power AR1}
\begin{tabular}{llcccc}
\hline
&method&$m=1$&$m=2$&$m=4$&$m=8$\\
\hline
$\phi=-0.1$& SIP 1&0.991&0.999&1.000&1.000\\
 & SIP 2&0.989&0.999&1.000&1.000\\
 & oracle&1.000&1.000&1.000&1.000\\
\hline
$\phi=-0.05$& SIP 1&0.557&0.678&0.728&0.653\\
 & SIP 2&0.529&0.651&0.716&0.670\\
 & oracle&0.999&0.997&0.986&0.974\\
\hline
$\phi=-0.025$& SIP 1&0.204&0.226&0.216&0.222\\
 & SIP 2&0.174&0.216&0.201&0.173\\
 & oracle&0.713&0.628&0.484&0.375\\
\hline
$\phi=0.025$& SIP 1&0.121&0.166&0.185&0.180\\
 & SIP 2&0.143&0.184&0.199&0.160\\
 & oracle&0.706&0.593&0.489&0.372\\
\hline
$\phi=0.05$& SIP 1&0.415&0.557&0.626&0.571\\
 & SIP 2&0.458&0.595&0.659&0.600\\
 & oracle&1.000&0.998&0.988&0.967\\
\hline
$\phi=0.1$& SIP 1&0.923&0.984&0.999&0.998\\
 & SIP 2&0.935&0.990&1.000&0.998\\
 & oracle&1.000&1.000&1.000&1.000\\
\hline
\end{tabular}
\end{center}
\end{table}
We observe that the SIP tests reliably uncover nontrivial autocorrelation unless it is extremely weak. Specifically, the SIP tests achieve power close to 1 for $\omega=\pm 0.1$ in the MA(1) case and $\phi=\pm 0.1$ in the AR(1) case. For MA(4) noise, we showcase three distinct patterns. In scenario 1, the SIP tests perform poorly with improperly chosen $m$. In scenario 2, the SIP tests outperform the oracle procedure, which suffers at $m=1$ due to weak lag-1 correlation. In scenario 3, all methods fail at $m=1$, but the SIP tests outperform the oracle when $m=2$. The oracle procedure lacks power when autocorrelations are weak and $m$ is too small because it discards covariance information beyond lag $m$. In contrast, in SIP tests, a smaller $m$ does not necessarily lead to power loss as shown in scenarios 2 and 3, since higher-lag covariance is implicitly incorporated into the test procedure.

In summary, we recommend $m=4$ as the default choice for both SIP tests. We slightly favor the SIP 2 test for larger $m$ due to its more accurate control of type I error.

\subsection{Real Data Analysis}

We apply the SIP tests to synthesized RNA nanopore sequencing reads from \cite{wang2024adapting}. The dataset comprises 900 sequences of integers representing ionic current levels, with an average sequence length exceeding 30,000. Figure \ref{fig1} in the introduction illustrates two examples of nanopore sequencing data, showcasing frequent mean shifts. While some of these shifts are visually apparent, accurately identifying the exact locations of all change points remains challenging.

\begin{figure}[h]
    \centering
    \includegraphics[width=14cm]{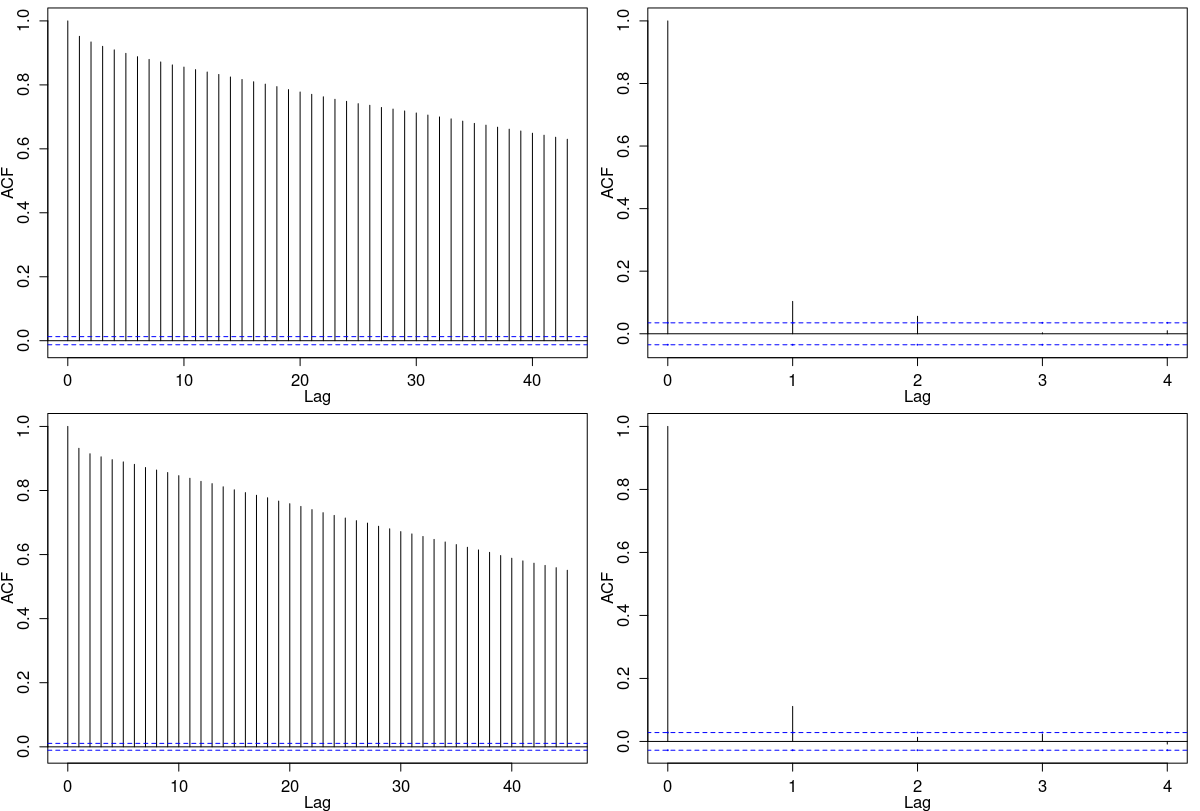}
    \caption{Left: the standard ACF plots for nanopore sequences with id$=33$ and 39; right: shift-immune ACF plots for nanopore sequences with id=33 and 39.}
    \label{fig2}
\end{figure}

We apply both SIP tests with $m=4$ to the 900 sequences, and the results indicate significant serial correlation in all sequences at the 5\% level. A summary of the $p$-values for both SIP tests is provided in Table \ref{nanopore test}. This confirms significant autocorrelation in nanopore sequencing data. To gain further insight into the autocorrelation structure, we visualize the low-order autocorrelations using our proposed SIP-ACF plot on the data shown in Figure \ref{fig1}, where the two sequences are from the data in \cite{wang2024adapting} labeled as 33 and 39, with UUIDs \texttt{read\_0abeb32d-c351-427b-ae41-ddf46ba1302d} and \texttt{read\_0c538684-1ea6-4f26-99ca-f36f5e278ea4}. The SIP-ACF plots, presented in the right column of Figure \ref{fig2}, reveal small but significant positive lag-1 autocorrelations. Similar patterns are observed in most of the nanopore sequences examined. We also detect small lag-2 to lag-4 autocorrelations, with significance levels varying among sequences. In contrast, the classical ACF plots, shown in the left column of Figure \ref{fig2}, completely fail to capture the autocorrelation structure due to the frequent mean shifts.

\begin{table}[ht]
\begin{center}
\caption{Summary of test results on nanopore sequencing data.}\label{nanopore test}
\begin{tabular}{lcc}
\hline
 &Average $p$-value&Maximum $p$-value\\

\hline
  SIP 1&1.408e-05&0.012\\
  SIP 2&1.032e-05&0.009\\
\hline
\end{tabular}
\end{center}
\end{table}

\section*{Appendix}
The theoretical results in Sections \ref{sec2} and \ref{sec:SIP} are derived in Appendices A and B, respectively.

\subsection*{Appendix A}

{\bf Proof of Proposition \ref{prop1}:} Simple calculation  shows
\begin{align*}
 \E\left(\bX^{\top} \bA \bX\right)
    &= \E\left[(\btheta+\bveps)^{\top} \bA (\btheta+\bveps)\right] \\
    &=  \E\left(\btheta^{\top} \bA \btheta +\btheta^{\top} \bA \bveps +\bveps^{\top} \bA \btheta +\bveps^{\top} \bA \bveps\right) \\
    &=  \btheta^{\top}\bA\btheta + \E\left(\bveps^{\top} \bA \bveps\right) \\
    &=\sum_{h=1-n}^{n-1} (n-|h|)a_{|h|}\gamma_h+ \btheta^{\top}\bA\btheta.
\end{align*}\hfill$\Box$

{\bf Proof of Theorem \ref{thm1}}: In the trivial case $L=1$, $\Theta_1=\mathbb{R}^n$. The quadratic form $\btheta^{\top} \bA\btheta = 0$ for all $\btheta$ if and only if $\bA=\bzero$, i.e. $a_i=0$ for all $0\leq i<n$. It is easy to check that the equations \eqref{V8}-\eqref{V11} hold if and only if $a_i=0$ for all $0\leq i<n$. Next we work on the more interesting cases.

We prove the necessity first. We are going to plug various $\btheta\in\Theta_L$ in the equality $\btheta^{\top} \bA\btheta = 0$ to achieve the equations \eqref{V8}-\eqref{V11}. Let $\bone_t$ and $\bzero_t$ be the $t$-dimensional vectors $(1,\dots,1)^{\top}$ and $(0,\dots,0)^{\top}$ respectively.
Let $\btheta= (\bone_L^{\top},\bzero_{n-L}^{\top})^{\top}$. $\btheta^{\top} \bA \btheta$ calculates the sum of all entries of the top-left $L\times L$ submatrix of $\bA$. So the equality $\btheta^{\top} \bA \btheta=0$ implies 
\begin{equation*}
    La_0 + 2(L-1)a_1 + \cdots + 4a_{L-2} +2a_{L-1}=0.
\end{equation*}
With $\btheta= (\bone_{L+1}^{\top},\bzero_{n-L-1}^{\top})^{\top}$, the equality $\btheta^{\top} \bA \btheta=0$ implies
\begin{equation} \label{proof prop1 2}
(L+1)a_0 + 2La_1 + \cdots + 4a_{L-1}+2a_{L}=0.
\end{equation}
It yields \eqref{V8} by taking the difference of \eqref{proof prop1 2} and its preceding equation. Note that both $(\bone_L^{\top},\bzero_{n-L}^{\top})^{\top}$ and $(\bone_{L+1}^{\top},\bzero_{n-L-1}^{\top})^{\top}$ are in $\Theta_L$ by the condition $L<n/2$. Now if we multiply \eqref{V8} by $(L+1)$ and then subtract \eqref{proof prop1 2} from it, we get \eqref{V9}. The equation \eqref{V10} is not necessary when $n=2L+1$. When $n\geq 2L+2$,
$(\bone_{L+2}^{\top},\bzero_{n-L-2}^{\top})^{\top}\in\Theta_L$. We can continue the process and get
\begin{equation} \label{proof prop1 4}
(L+2)a_0 + 2(L+1)a_1 + \cdots + 4a_{L}+2a_{L+1}=0.
\end{equation}
Subtracting the sum of \eqref{proof prop1 2} and \eqref{V8} from \eqref{proof prop1 4}, we obtain $a_{L+1}=0$. Following the same procedure with $\btheta=(\bone_{L+3}^{\top},\bzero_{n-L-3}^{\top})^{\top},\,\dots,\, (\bone_{n-L}^{\top},\bzero_{L}^{\top})^{\top}$, we can derive $a_{L+1}= \cdots =a_{n-L-1}=0$. Therefore, \eqref{V10} holds.

Finally, we employ $\btheta=\bone_n$ in the equality $\btheta^{\top} \bA \btheta=0$ and obtain
\begin{align*}
na_0 + 2(n-1)a_1 + \cdots + 4a_{n-2}+2a_{n-1}=0
\end{align*}
which yields \eqref{V11} by applying the results  \eqref{V8}-\eqref{V10}.

Now we show the sufficiency. For any $\btheta\in\Theta_L$, as in \eqref{V2}, we can write $\btheta=\sum_{j=1}^{J+1}\mu_j\btheta_j$ where
$\btheta_1=(\bone_{\tau_1}^{\top},\bzero_{n-\tau_1}^{\top})^{\top}$, $\btheta_j=(\bzero_{\tau_j}^{\top},\bone_{\tau_{j+1}-\tau_j}^{\top},\bzero_{n-\tau_{j+1}}^{\top})^{\top}$,
for $j=2,\dots,J$, and $\btheta_{J+1}=(\bzero_{\tau_J}^{\top},\bone_{n-\tau_J}^{\top})^{\top}$.
\begin{equation*}
    \btheta^{\top} \bA \btheta = \sum_{j=1}^{J+1}\sum_{\ell=1}^{J+1} \mu_j\mu_{\ell} \btheta_j^{\top}\bA\btheta_{\ell}.
\end{equation*}
It suffices to show  $\btheta_j^{\top}\bA\btheta_{\ell}=0$ for all $j$, $\ell$. It is straightforward to see $(\sum_{\ell=1}^j\btheta_{\ell})^{\top}\bA(\sum_{\ell=1}^j\btheta_{\ell})=0$ for $1\leq j\leq J+1$ from the proof of the necessity part. In particular, we have $\btheta_1^{\top}\bA\btheta_1=0$ and $(\btheta_1+\btheta_2)^{\top}\bA(\btheta_1+\btheta_2)=0$, where the latter gives  $\btheta_1^{\top}\bA\btheta_1+2\btheta_1^{\top}\bA\btheta_2+\btheta_2^{\top}\bA\btheta_2=0$. Note that $\btheta_2^{\top}\bA\btheta_2=0$ as $\bA$ is Toeplitz. (The fact that $\bA$ is Toeplitz implies that $\btheta_2^{\top}\bA\btheta_2$ won't change if we move all the 1s in $\btheta_2$ to the top coordinates.) Therefore, we can claim $\btheta_1^{\top}\bA\btheta_2=0$. By the same argument, we can get  $\btheta_j^{\top}\bA\btheta_{j+1}=0$. Moreover, it follows \eqref{V10} that $\btheta_j^{\top}\bA\btheta_{\ell}=0$ for $\ell-j>1$. \hfill $\Box$

{\bf Proof of Proposition \ref{prop:eve}}: The quadratic form $\bX^\top\bA\bX$ described in Theorem \ref{thm1} is invariant under a global mean shift if and only if $\bA\bone=\bzero$, i.e., the row sums of $\bA$ are all zero. The sufficiency in Proposition~\ref{prop:eve} is then obvious due to equation \eqref{V8}. To show the necessity, we first observe that the sum of $(L+1)$-th row of $\bA$, given equation \eqref{V10}, is
\begin{equation*}
    a_0+2a_1+\cdots+2a_{L-1}+a_L+a_{n-L}=0,
\end{equation*}
which implies $a_{n-L}=a_L$ in view of \eqref{V8}. Next we move up to the $L$-th row:
\begin{equation*}
    a_0+2a_1+\cdots+2a_{L-2}+a_{L-1}+a_L+a_{n-L}+a_{n-L+1}=0,
\end{equation*}
which, together with \eqref{V8} and $a_{n-L}=a_L$, implies that $a_{n-L+1}=a_{L-1}$. The proof of necessity is completed by moving upward of the equation $\bA\bone=\bzero$ row by row till the top. \hfill $\Box$

{\bf Proof of Proposition \ref{prop:T rep}}: We can express $T_h$ as
\begin{equation}\label{T_h}
  T_h=\sum_{i=1}^{n}(X_i-X_{i+h})^2 = 2\sum^n_{i=1} X_i^2 - 2\sum^n_{i=1} X_iX_{i+h}, \quad h=1,\dots, L.
\end{equation}
That is, each $T_h$ can be rendered as a quadratic form using the circulant matrix induced by $(2, 0,\ldots,0,-1,0,\ldots, 0,-1,0,\ldots,0)$, where the two $-1$ appear as the $(h+1)$-th and $(n-h+1)$-th entries.
It is straightforward to verify \eqref{eq:XAX_T} with equation \eqref{V8}. \hfill $\Box$

{\bf Proof of Proposition \ref{prop:XAX_autocov}}: Since $\bA\in\cA_L$, it holds that $\btheta^\top\bA\btheta=0$. The proposition is then an immediate consequence of Proposition~\ref{prop1}.\hfill $\Box$

\subsection*{Appendix B}

{\bf Proof of Proposition \ref{Prop3}}: Since $\{\varepsilon_i\}$ is $m$-dependent, by Proposition~\ref{prop1} and Theorem~\ref{thm1}
\begin{equation*}
    \E \hat\gamma_h = 2(n-h)a_{hh}\gamma_h = (1-h/n)\gamma_h.
\end{equation*}
According to \eqref{eq:XAX_T}, the $\hat\gamma_0$ defined in \eqref{gamma0 hat} can be represented as $\bX^\top\bA_0\bX$, where $\bA_0$ takes the form \eqref{V6} with the first row
\begin{equation*}
    \left(1/n,0,\ldots,0,-\frac{m+2}{2n}, \frac{m+1}{2n},0,\ldots,0,\frac{m+1}{2n},-\frac{m+2}{2n},0,\ldots,0\right),
\end{equation*}
where the five nonzero values are at the 1st, $(m+2)$-th, $(m+3)$-th, $(n-(m+2))$-th, and $(n-(m+1))$-th entries. The expectation of $\hat\gamma_0$ is therefore $\gamma_0$ again by Proposition~\ref{prop1}.
\hfill $\Box$

We shall defer the proof of Proposition~\ref{prop:PT} after that of Theorem~\ref{thm2}, as it could make use of some arguments in the latter.
The proof of Theorem \ref{thm2} relies on the following lemma which  establishes the asymptotic property of $\{T_h\}$ as defined in \eqref{T_h}.
\begin{lemma} \label{lemma1}
Under conditions 1, 2, and 3, and the null hypothesis that $\eps_1, \dots, \, \eps_n$ are IID, if $K\leq L/2$, then $\bT_K=(T_1,...,T_K)^{\top}$ is asymptotically normal as $n\to\infty$,
\begin{equation}
\label{eq:lem1_clt}
  \bSigma_{K,w}^{-1/2}\sqrt{n}(\bT_K/n- \bnu_K )\xrightarrow{D} \cN(\bzero_K, \bI_K),
\end{equation}
where
\begin{equation*}
    \bSigma_{K,w}= 4 \gamma_0^2 \left[\bI_K+(\kappa_4-1)\bone_K\bone_K^{\top}+2w\bH_K\right],\quad
    \bnu_K=\gamma_0(2\cdot\bone_K+ w\bEta_K).
\end{equation*}
Moreover, the range of $K$ can be relaxed to $K\leq L$, in which case
\begin{equation}
\label{eq:sto_dom}
  \lim_{n\rightarrow\infty}\bSigma_{K,2w}^{-1/2}\sqrt{n}(\bT_K/n- \bnu_K )\preceq \cN(\bzero_K, \bI_K).
\end{equation}

\end{lemma}

{\bf Proof of Lemma \ref{lemma1}.} By Proposition 2.1 in \cite{hao2023equivariant}, the expectation and covariance of $n^{-1}\bT_K$ are $\bnu_K$ and $n^{-1}\bSigma_{K,w}$, respectively, under the null hypothesis. In particular, the variance of $T_k$ is $4n\gamma_0^2(\kappa_4+2kw)$. For ease of presentation, we will prove a central limit theorem for $(2\gamma_0)^{-1}\cdot[n(\kappa_4+2kw)]^{-1/2}\cdot (T_k-n\nu_k)$ for every $1\leq k\leq K$. The central limit theorem for $\bT_K$ can be established along similar lines with an application of the Cram\'{e}r-Wold device.

Write $T_k$ as
\begin{align*}
  T_k&=\sum_{i=1}^{n}(X_i-X_{i+k})^2\\
  &=\sum_{i=1}^{n}(\theta_i-\theta_{i+k}+\v_i-\v_{i+k})^2\\
  &=\sum_{i=1}^{n}(\theta_i-\theta_{i+k})^2+2\sum_{i=1}^{n}(\theta_i-\theta_{i+k})(\v_i-\v_{i+k})+\sum_{i=1}^{n}(\v_i-\v_{i+k})^2\\
  &=n(2\gamma_0+kw)+2\sum_{i=1}^{n}(\theta_i-\theta_{i+k})(\v_i-\v_{i+k})+\sum_{i=1}^{n}\left[(\v_i-\v_{i+k})^2-2\gamma_0\right] \\
  & = n(2\gamma_0+kw)+2\sum_{i=1}^{n} d_{ki}\v_i + 2\sum_{i=1}^{n} (\v_i^2-\gamma_0) - 2\sum_{i=1}^{n}\v_i\v_{i+k},
\end{align*}
where all the indices are modulo $n$, and $d_{ki}=\mu_{j}-\mu_{j+1}$ for $\tau_j-k<i\leq\tau_j$, $d_{ki}=\mu_{j+1}-\mu_{j}$ for $\tau_j<i\leq\tau_j+k$, and $d_{ki}=0$ otherwise.
We shall apply the martingale central limit theorem (see Corollary~3.1 of \cite{hall1980martingale}). Note that since the number, location and magnitude of the change-points are allowed to change with the sample size, we are using the triangular-array version of the martingale and its central limit theorem. In order to have a more homogeneous form of the martingale differences, we will use indices that are no longer modulo $n$ for the rest of this proof, and modify $T_k$ slightly by considering
\begin{align}
\label{eq:T_tilde}
    \tilde T_k = n(2\gamma_0+kw)+2\sum_{i=1}^{n} d_{ki}\v_i + 2\sum_{i=1}^{n} (\v_i^2-\gamma_0) - 2\sum_{i=1}^{n}\v_i\v_{i-k},
\end{align}
where $\v_{0},\v_{-1},\dots,\v_{1-k}$ are IID with other $\v_i$'s. It is clear that $\tilde T_k-T_k = O_P(1)$, so it suffices to have a central limit theorem for $\tilde T_k$. Let
\begin{equation*}
    D_{ni} := (2\gamma_0)^{-1}\cdot[n(\kappa_4+2kw)]^{-1/2}[2d_{ki}\v_i+ 2(\v_i^2-\gamma_0) - \v_i\v_{i-k}],
\end{equation*}
and $\mathcal F_i$ be the $\sigma$-field generated by $\{\v_{1-k},\ldots,\v_0,\v_1,\v_2,\ldots,\v_i\}$, then $\{D_{ni}\}$ is a martingale difference sequence with respect to the filtration $\{\mathcal F_i\}$. To apply the martingale central limit theorem, it suffices to show that
\begin{align*}
    \sum_{i=1}^n \E\left\{ D_{ni}^2I[|D_{ni}|\geq \epsilon]\right\} \rightarrow 0, \quad\forall\, \epsilon>0.
\end{align*}
Let $u_i:=4\v_i^2+4(\v_i^2-\gamma_0)^2+\v_i^2\v_{i-k}^2$, then the preceding sum is upper bounded by
\begin{align*}
    & \left[4n\gamma_0^2(\kappa_4+2kw)\right]^{-1}\sum_{i=1}^n \E\left\{(d_{ki}^2+2)u_iI\left[|u_i|\geq \frac{\epsilon^2\cdot 4n\gamma_0^2(\kappa_4+2kw)}{(d_{ki}^2+2)}\right]\right\} \\
    \leq & \frac{2n(1+\gamma_0wk)}{4n\gamma_0^2(\kappa_4+2kw)} \E\left\{u_iI\left[|u_i|\geq \frac{\epsilon^2\cdot 4n\gamma_0^2(\kappa_4+2kw)}{(\max_i d_{ki}^2+2)}\right]\right\}.
\end{align*}
Condition~3 entails that $\max_i d_{ki}^2= o(n)$, so the expectation in the preceding equation goes to zero as $n\rightarrow\infty$, and the proof of \eqref{eq:lem1_clt} is complete.

The second statement \eqref{eq:sto_dom} for the case $K\leq L$ follows the fact that the covariance matrix of $\bT_K/\sqrt{n}$ is upper bounded by $\bSigma_{K,2w}$, see the proof of the second part of Theorem 2.4 in \cite{hao2023equivariant}. \hfill $\Box$

{\bf Proof of Theorem \ref{thm2}}:
As $\hat{\gamma}_0$ is the linear combination of $T_{m+1}$ and $T_{m+2}$,
whose asymptotic joint distribution is shown in Lemma \ref{lemma1}, it is straightforward to show the asymptotic distribution of $\hat{\gamma}_0$ as
\begin{equation*}
    \frac{\sqrt{n}(\hat{\gamma}_0 - \gamma_0)}{\gamma_0\sqrt{\left[\kappa_4+2(m+1)(m+2)(1+w)\right]}} \to \cN\left(0,  1 \right),
\end{equation*}
Note that Condition 3 implies $w=W(\btheta)/(n\gamma_0)\leq n\cdot  \max_j\{(\mu_j-\mu_{j+1})^2\}/(n\gamma_0)=o(n)$. Together with the finite fourth moment condition, this implies that the variance of $\hat{\gamma}_0$ approaches zero as $n\to\infty$.
Therefore we can conclude that $\hat{\gamma}_0$ converges in probability to $\gamma_0$ by Chebyshev's inequality.

Similarly, the asymptotic normality of $\hat{\bgamma}$ is implied by \eqref{gamma_hat1}, \eqref{gamma0 hat} and Lemma \ref{lemma1}, as $\hat{\bgamma}$ is a linear function of $\bT_{m+2}$. In particular,  $\hat{\bgamma}=\bR\bT_{m+2}/(2n)$, where
\[\bR = \left(
  \begin{array}{cccccc}
    -1 & 0 & \cdots & 0 & m+1  & -m \\
    0 & -1 & \cdots & 0 & m  & -(m-1) \\
    \vdots & \vdots & \ddots & \vdots & \vdots &\vdots  \\
    0 & 0 & \cdots & -1 & 2 &-1 \\
  \end{array}
\right) = \left(-\bI_m,(m+2)\bone_m- \bEta_m,\bEta_m-(m+1)\bone_m\right).
\]
By Lemma \ref{lemma1}, the covariance matrix $\bSigma_{m+2,w}$ of $\bT_{m+2}/n$ could be decomposed as \[\bSigma_{m+2,w} = 4n^{-1}\gamma_0^2\left(
    \begin{array}{ccc}
    \bSigma_m/(4n^{-1}\gamma_0^2) & (\kappa_4-1)\bone_m+2w\bEta_m & (\kappa_4-1)\bone_m+2w\bEta_m \\
    (\kappa_4-1)\bone_m^{\top}+2w\bEta_m^{\top} & \kappa_4+2(m+1)w & (\kappa_4-1)+2(m+1)w \\
    (\kappa_4-1)\bone_m^{\top}+2w\bEta_m^{\top} & (\kappa_4-1)+2(m+1)w & \kappa_4+2(m+2)w\\
  \end{array}
\right).\]
Therefore, $\bSigma_{\gamma} =\bR\bSigma_{m+2,w}\bR^{\top}/4$, which is further detailed as
\begin{align*}
  \bSigma_{\gamma} =&\bR\bSigma_{m+2,w}\bR^{\top}/4\\
  =& \gamma_0^2\{\bI_m+\left[(2m^2+6m+5)+2(m^2+3m+2)w\right]\bone_m\bone_m^{\top}\\
   \quad & -\left[(2m+3)+2(m+2)w\right](\boldsymbol{\eta}_{m}\bone_{m}^{\top} +\bone_{m}\boldsymbol{\eta}^{\top}_{m})\\&+(2+2w)\boldsymbol{\eta}_{m}\boldsymbol{\eta}^{\top}_{m} +2w\bH_m\}.
\end{align*}
The proof is complete. \hfill $\Box$

\noindent {\bf Remark 1.} We would like to point out that while the kurtosis $\kappa_4$ of $\varepsilon_i$ appears in the covariance matrix $\bSigma_{m+2,w}$ of $\bT_{m+2}$, it no longer shows up in the covariance matrix $\bSigma_\gamma$ of $\hat\bgamma$. This is because all the square terms $\varepsilon_i^2$ are eliminated from $\hat\bgamma$ in order to remove $\gamma_0$ from its expectation. As a result, the fourth moment does not appear in the variance calculation.

{\bf Proof of Proposition \ref{prop:PT}}:
We divide the proof into two steps: first
\begin{equation}
\label{eq:PT_expectation}
    \lim_{n\rightarrow\infty}\cP_{m+2}[\E\bT_{m+2}] = \cP_{m+2}\bgamma_{m+2},
\end{equation}
and next,
\begin{equation}
\label{eq:PTconv}
    \cP_{m+2}\bT_{m+2} -\cP_{m+2}[\E\bT_{m+2}] \xrightarrow{P} \bzero.
\end{equation}
A direct calculation shows that $\E [T_h/(2n)] = \gamma_0 + hw/2 + (1-h/n)\gamma_h + (h/n)\gamma_{n-h}$ for $1\leq h\leq (m+2)$. According to Theorem~\ref{thm1} and Proposition~\ref{prop:XAX_autocov}, the expectation of $\cP_{m+2}\bT_{m+2}$ does not involve $\btheta$ and $\gamma_0$, and hence \eqref{eq:PT_expectation} follows.
To show \eqref{eq:PTconv}, it suffices to show it for $\tilde \bT_{m+2}$ instead of $\bT_{m+2}$, where $\tilde\bT_{m+2}:=(\tilde T_1,\tilde T_2,\ldots,\tilde T_{m+2})^\top$, and $\tilde T_k$ is defined in \eqref{eq:T_tilde}. Again by Theorem~\ref{thm1} and Proposition~\ref{prop:XAX_autocov}, the terms $\gamma_0$, $w$ and $\varepsilon_i^2$ will all disappear in $\cP_{m+2}\tilde \bT_{m+2}$, so it in turn suffices to show that
\begin{equation*}
    n^{-1}\sum_{i=1}^{n} d_{ki}\v_i - n^{-1}\sum_{i=1}^{n}(\v_i\v_{i-k}-\gamma_k) \xrightarrow{P} 0
\end{equation*}
for every $1\leq k\leq m+2$. The convergence of the second term $n^{-1}\sum_{i=1}^{n}(\v_i\v_{i-k}-\gamma_k)$ to zero in probability is warranted by the ergodic theorem, under Condition~2. The first term converges to zero in probability due to Condition~3. The proof is complete.
\hfill $\Box$

The following lemma establishes the consistency of $\hat w_1$ and $\hat w_2$, which plays a key role in the proof of Theorem \ref{thm3}.
\begin{lemma}\label{lemma2}
(i) Under conditions 1-3, and $\{\varepsilon_i\}$ is $m$-dependent with $m+2\leq L$, $\hat w_1- w\xrightarrow{P} 0$.

(ii) Under one more condition that $\eps_1, \dots, \, \eps_n$ are IID, $\hat w_2- w\xrightarrow{P} 0$.
\end{lemma}

{\bf Proof of Lemma \ref{lemma2}}:
Part (i). When $h>m$,   $E(X_iX_{i+h})=E(X_i)E(X_{i+h})=\theta_i\theta_{i+h}$. It follows that
\begin{align*}
 &E \left(\sum^{n}_{i=1}X_i X_{i+m+2}-\sum^{n}_{i=1}X_i X_{i+m+1}\right)\\
 =&  \sum^{n}_{i=1}\theta_i \theta_{i+m+2}-\sum^{n}_{i=1}\theta_i \theta_{i+m+1} \\
 =& \frac12\sum^{n}_{i=1}(\theta_i-\theta_{i+m+2})^2-\frac12\sum^{n}_{i=1}(\theta_i-\theta_{i+m+1})^2\\
 =&\frac{m+2}{2}W(\btheta)-\frac{m+1}{2}W(\btheta)\\
 =&\frac12W(\btheta)\\
 =&n\gamma_0w/2.
\end{align*}
Moreover,
\begin{align*}
    &\Var\left(\sum^{n}_{i=1}X_i X_{i+m+2}-\sum^{n}_{i=1}X_i X_{i+m+1}\right)\\
    =&\Var\sum_{i=1}^n  \left(\theta_i \varepsilon_{i+m+2}+\theta_{i+m+2}\varepsilon_i +\varepsilon_i \varepsilon_{i+m+2} -\theta_i \varepsilon_{i+m+1}-\theta_{i+m+1}\varepsilon_i -\varepsilon_i \varepsilon_{i+m+1}\right)\\
    =&\Var\left(\sum_{i=1}^n  (\theta_i-\theta_{i+1})  \varepsilon_{i+m+2}+\sum_{i=1}^n (\theta_{i+m+2}-\theta_{i+m+1})\varepsilon_i +\sum_{i=1}^n\varepsilon_i \varepsilon_{i+m+2}    -\sum_{i=1}^n\varepsilon_i \varepsilon_{i+m+1}\right)\\
    \leq&4\Var \sum_{i=1}^n  (\theta_i-\theta_{i+1})  \varepsilon_{i+m+2}+4\Var\sum_{i=1}^n (\theta_{i+m+2}-\theta_{i+m+1})\varepsilon_i +4\Var\sum_{i=1}^n\varepsilon_i \varepsilon_{i+m+2}    +4\Var\sum_{i=1}^n\varepsilon_i \varepsilon_{i+m+1}\\
    \leq & 8W(\btheta)\gamma_0+ 8n(2m+1)\gamma_0^2.
\end{align*}
In the last inequality, we use two facts. First, $\sum_{i=1}^n (\theta_i-\theta_{i+1})  \varepsilon_{i+m+2}$ is a linear combination of at most $J+1$ IID random variables because of the $m$-dependent assumption, $\btheta\in\Theta_L$ and $L>m$. As a result,  $\Var\sum_{i=1}^n (\theta_i-\theta_{i+1})  \varepsilon_{i+m+2}\leq \gamma_0\sum_{i=1}^n (\theta_i-\theta_{i+1})^2=\gamma_0W(\btheta)$. Similarly, we can bound $\Var\sum_{i=1}^n (\theta_{i+m+2}-\theta_{i+m+1})\varepsilon_i$.
Second, to bound \[\Var\sum_{i=1}^n\varepsilon_i \varepsilon_{i+m+1}=\Cov\left(\sum_{i=1}^n\varepsilon_i \varepsilon_{i+m+1},\sum_{j=1}^n\varepsilon_j \varepsilon_{j+m+1}\right),\]
we observe that for each $i$, there are at most $2m+1$ different $j$ $(i-m\leq j\leq i+m)$ such that
$\Cov\left( \varepsilon_i \varepsilon_{i+m+1},\varepsilon_j \varepsilon_{j+m+1}\right)\ne0$. For those pairs, by the Cauchy-Schwarz inequality,
\[\Cov\left( \varepsilon_i \varepsilon_{i+m+1},\varepsilon_j \varepsilon_{j+m+1}\right)\leq \sqrt{\Var (\varepsilon_i \varepsilon_{i+m+1})\Var(\varepsilon_j \varepsilon_{j+m+1})}=\gamma_0^2.\]
Therefore, we have
\[\Var\sum_{i=1}^n\varepsilon_i \varepsilon_{i+m+2} \leq n(2m+1)\gamma_0^2.\]

Given the calculations of expectation and variance above, we conclude
\[n^{-1}\left(\sum^{n}_{i=1}X_i X_{i+m+2}-\sum^{n}_{i=1}X_i X_{i+m+1}\right)\] is an unbiased estimator of $\gamma_0w/2$ with variance at most \[(8W(\btheta)\gamma_0+ 8n(1+2m)\gamma_0^2)/n^2=8\gamma_0^2(w+1+2m)/n,\] which goes to 0 as $n\to\infty$. As a result,
\[2n^{-1}\left(\sum^{n}_{i=1}X_i X_{i+m+2}-\sum^{n}_{i=1}X_i X_{i+m+1}\right)-\gamma_0w \xrightarrow{P} 0,\] and
\[2(n\hat\gamma_0)^{-1}\left(\sum^{n}_{i=1}X_i X_{i+m+2}-\sum^{n}_{i=1}X_i X_{i+m+1}\right)-w \xrightarrow{P} 0,\]
for any consistent estimator $\hat\gamma_0$.

Part (ii). The property of the least squares estimator $(\hat{\alpha},\hat\beta)^{\top}$ is studied in \cite{hao2023equivariant} where Theorem 2.1 shows that it is unbiased with asymptotic variance $n^{-1}O(1+w)$. Note that Condition 3 implies $w=W(\btheta)/(n\gamma_0)\leq n\cdot  \max_j\{(\mu_j-\mu_{j+1})^2\}/(n\gamma_0)=o(n)$. Therefore, $(\hat{\alpha},\hat\beta)^{\top}$ is unbiased and its variance goes to 0, which implies $(\hat{\alpha},\hat\beta)^{\top}\xrightarrow{P} (\alpha,\beta)^{\top}$. By definition $\alpha=\gamma_0>0$, so $\hat w_2$ is consistent.\hfill $\Box$

{\bf Proof of Theorem \ref{thm3}}: For any autocorrelation estimator $\hat\brho=\hat\bgamma/\hat{\gamma}_0$ where $\hat{\gamma}_0$ is consistent, it follows from Theorem \ref{thm2} and Slutsky's theorem that  $\bSigma_{\rho,w}^{-\frac12} \sqrt{n}\hat{\brho}  \xrightarrow{D} \cN(\bzero_{m}, \bI_m)$, where
$\bSigma_{\rho,w}=\bSigma_{\gamma,w}/\gamma_0^2$.

Now we show $ {\bSigma}_{\rho,\hat w}^{-\frac12} \bSigma_{\rho,w}^{\frac{1}{2}}\xrightarrow{P}\bI_m$. We can write ${\bSigma}_{\rho,\hat w}^{-\frac12} \bSigma_{\rho,w}^{\frac{1}{2}} = (\bM_1+\hat w \bM_2)^{-\frac12}(\bM_1+w\bM_2)^{-\frac12}$, where
$$\bM_1=\bI_m+(2m^2+6m+5)\bone_m\bone_m^{\top}-(2m+3)(\bEta_{m}\bone_{m}^{\top} +\bone_{m}\bEta^{\top}_{m})+2\bEta_m\bEta_m^{\top},$$
$$\bM_2= 2(m^2+3m+2)\bone_m\bone_m^{\top}-2(m+2)(\bEta_{m}\bone_{m}^{\top} +\bone_{m}\bEta^{\top}_{m})+2\bEta_m\bEta_m^{\top}+2\bH_m$$
are both positive definite. To see $\bM_2$ is positive definite, note that $\bM_2=2\bR\bH_{m+2}\bR^{\top}$ from the proof of Theorem \ref{thm2}. The Cholesky decomposition gives $\bH_{m+2}=\bC\bC^{\top}$ where $\bC$ is a lower triangular matrix with $C_{ij}=1$ for all $i\leq j$. It implies $\bH_{m+2}$ is positive definite. Moreover, $\bR$ is of full rank as it contains $-\bI_m$. Overall, this indicates $\bM_2$ is positive definite. For $\bM_1$, the proof is similar but easier, so we omit the details here.

Lemma \ref{lemma2} implies that
\begin{align*}
 {\bSigma}_{\rho,\hat w}^{-\frac12} \bSigma_{\rho,w}^{\frac{1}{2}} &= (\bM_1+ w \bM_2 +(\hat w-w)\bM_2)^{-\frac12}(\bM_1+w\bM_2)^{-\frac12}\\
 &=(\bM_1+ w \bM_2 +o_P(1))^{-\frac12}(\bM_1+w\bM_2)^{-\frac12}\\
 &=((\bM_1+ w \bM_2 )^{-\frac12}+o_P(1))(\bM_1+w\bM_2)^{-\frac12}\\
 &=\bI_m+o_P(1).
\end{align*}

By Slutsky's theorem, $ {\bSigma}_{\rho,\hat w}^{-\frac12} \sqrt{n}\hat{\brho}=\left({\bSigma}_{\rho,\hat w}^{-\frac12} \bSigma_{\rho,w}^{\frac{1}{2}}\right)\left( \bSigma_{\rho,w}^{-\frac{1}{2}}\sqrt{n}\hat{\brho}\right)\xrightarrow{D} \cN(\bzero_{m}, \bI_m)$, and \eqref{chi2} holds.\hfill $\Box$

{\bf Proof of Proposition \ref{prop5}}: Direct calculation of the right bottom corner of the matrix leads to
\[1+ [(2m^2+6m+5)+2(m^2+3m+2)\hat{w} ] \cdot 1
    - [(2m+3)+2(m+2)\hat{w} ](2m)+(2+2\hat{w})m^2 +2\hat{w}m,\]
which equals to $6+4\hat w$.\hfill $\Box$

\section*{Funding}
The authors are partially supported by the National Science Foundation, the University of Arizona Health Sciences Career Development Award and the University of Arizona Accelerate For Success Award.

\bibliography{changepointrefs2017}

\end{document}